\documentclass[prl,amsmath,amssymb,aps,superscriptaddress,floatfix,preprintnumbers, onecolumn,notitlepage,10pt,reprint]{revtex4-1}
\usepackage{times}
\usepackage{graphicx}
\usepackage{amsmath, braket,amsfonts}
\usepackage{amssymb}
\usepackage{natbib}
\usepackage{bm, color, ulem}
\usepackage{wrapfig}

\begin{document}
\renewcommand{\emph}{\textit}
\newcommand{\nocontentsline}[3]{}
\newcommand{\tocless}[2]{\bgroup\let\addcontentsline=\nocontentsline#1{#2}\egroup}

\title{Observation of fractional Chern insulators in a van der Waals heterostructure}
\author{E.M. Spanton$^*$}\affiliation{California Nanosystems Institute, University of California at Santa Barbara, Santa Barbara, CA, 93106}
\author{A.A. Zibrov$^*$}
\affiliation{Department of Physics, University of California, Santa Barbara CA 93106 USA}
\author{H. Zhou}\affiliation{Department of Physics, University of California, Santa Barbara CA 93106 USA}
\author{T. Taniguchi}\affiliation{Advanced Materials Laboratory, National Institute for Materials Science, Tsukuba, Ibaraki 305-0044, Japan}
\author{K. Watanabe}
\affiliation{Advanced Materials Laboratory, National Institute for Materials Science, Tsukuba, Ibaraki 305-0044, Japan}
\author{M. P. Zaletel}
\affiliation{Department of Physics, Princeton University, Princeton, NJ 08544, USA}
\author{A.F. Young}
\affiliation{Department of Physics, University of California, Santa Barbara CA 93106 USA}
\date{\today}%

\begin{abstract}
\textbf{  Topologically ordered phases are characterized by long-range quantum entanglement and fractional statistics rather than by symmetry breaking. First observed in a fractionally filled continuum Landau level, topological order has since been proposed to arise more generally at fractional filling of topologically non-trivial ``Chern'' bands. Here, we report the observation of gapped states at fractional filling of Harper-Hofstadter bands arising from the interplay of a magnetic field and a superlattice potential in a bilayer graphene/hexagonal boron nitride heterostructure. We observe new phases at fractional filling of bands with Chern indices $\mathcal{C} = -1, \pm 2,$ and $\pm 3$. Some of these, in $\mathcal{C}=-1$ and $\mathcal{C}=2$ bands, are characterized by fractional Hall conductance---they are `fractional Chern insulators' and constitute a new example of topological order beyond Landau levels.}
\end{abstract}

\maketitle
Band gaps in electronic systems can be classified by their symmetry and topology\cite{ryu_topological_2010}.
In two dimensions with no symmetries beyond charge conservation, for example, band gaps are classified by their Hall conductance, which takes  quantized integer values, $\sigma_{xy} = t \frac{e^2}{h}, t \in \mathbb{Z}$\cite{thouless_quantized_1982}.
Such integer quantum Hall (IQH) effects were first observed in isotropic two dimensional electron systems (2DES) subject to a large magnetic field\cite{klitzing_new_1980}.
These systems are very nearly translation invariant, in which case $t$ is fixed by the magnetic field $B$ and the electron density $n$, via $n = \frac{e}{h} t B$, with some  disorder required for the formation of plateaus in the Hall conductance\cite{girvin_quantum_1999}.  Recently, there has been interest in  systems where continuous translation invariance is strongly broken by a lattice, decoupling the Hall conductance from the magnetic field. A notable example is Haldane's staggered flux model\cite{Haldane}, which has non-zero quantized Hall conductance even when the net magnetic field is zero.
Bands which contribute a non-zero Hall conductance are called ``Chern'' bands, in reference to the underlying topological index of the band structure, the Chern number $\mathcal{C}$\cite{thouless_quantized_1982}, and
IQH effects resulting from filled Chern bands are known as ``Chern insulators'' (CI). A filled continuum Landau level (LL) is a special case of a CI, but more recently CIs in which $t$ is decoupled from $\frac{n}{B}$ have been observed in magnetically doped thin films with strong spin orbit interactions\cite{chang_experimental_2013} and in the Harper-Hofstadter\cite{thouless_quantized_1982} bands of graphene subjected to a superlattice potential \cite{Dean13Nat,Ponomarenko13Nat,Hunt13Sci}. The Haldane model has been engineered using ultracold atoms in an optical lattice\cite{jotzu_experimental_2014}.   

Interactions expand the topological classification of gapped states, allowing the Hall conductance $t$ to be quantized to a rational \emph{fraction}.
By Laughlin's flux-threading argument, an insulator with $t = \frac{p}{q}$ must have a fractionalized excitation with charge $\frac{e}{q}$\cite{laughlin_anomalous_1983}. A fractionally quantized Hall conductance in a bulk insulator is thus a smoking-gun signature of topological order, and fractional quantum Hall (FQH) effects have been observed in partially-filled continuum LLs in a variety of experimental systems\cite{TsuiStormerGossard82,du_fractional_2009,bolotin_observation_2009,tsukazaki_observation_2010}.  Can a ``fractional Chern insulator'' (FCI) arise from fractionally filling a more general Chern band? While a FQH effect in a LL may be considered a special case of a FCI, in this work we focus on FCIs which require a lattice for their existence.

The phenomenology of lattice FCIs differs from that of continuum LLs.
Chern bands with $\mathcal{C}\neq1$ can arise, leading to different ground states than are allowed in $\mathcal{C}=1$ LLs. In addition, unlike LLs, Chern bands generically have a finite, tunable bandwidth that competes with interactions, providing a new setting for the study of quantum phase transitions. Finally, FCIs might be found in experimental systems where Chern bands, but not LLs, are realizable.
A large body of theoretical work has begun to investigate these issues \cite{parameswaran_fractional_2013, bergholtzreview,sorenson,palmer,MollerCooper,Sheng,Neupert,Regnault}. 

Here, we report the experimental discovery of FCIs in a bilayer graphene (BLG) heterostructure at high magnetic field. The requirements to realize an FCI in an experimental system are, first, the existence of a Chern band, and, second, electron-electron interactions strong enough  to overcome both disorder and band dispersion.  We satisfy these requirements by using a high quality bilayer graphene heterostructure, in which the bilayer is encapsulated between hexagonal boron nitride (hBN) gate dielectrics and graphite top- and bottom gates (see Fig.~1A-B). This geometry was recently demonstrated to significantly decrease disorder, permitting the observation of delicate FQH states\cite{zibrov_tunable_2017}. We generate Chern bands by close rotational alignment ($\sim 1^{\circ}$) between the bilayer graphene and one of the two encapsulating hBN crystals.  Beating between the mismatched crystal lattices leads to a long-wavelength ($\sim$10 nm) moir\'e pattern that the electrons in the closest layer experience as a periodic superlattice potential (Fig.~1B(see Supplementary Information)). At high magnetic fields, the single particle spectrum of an electron in a periodic potential forms the Chern bands of the Hofstadter butterfly\cite{Dean13Nat,Ponomarenko13Nat,Hunt13Sci}.
These bands are formally equivalent to the Chern bands proposed to occur in zero-magnetic field lattice models\cite{parameswaran_fractional_2013}.

\begin{figure*}[t!]
\begin{center}
\includegraphics[page=1, width=6.9 in]{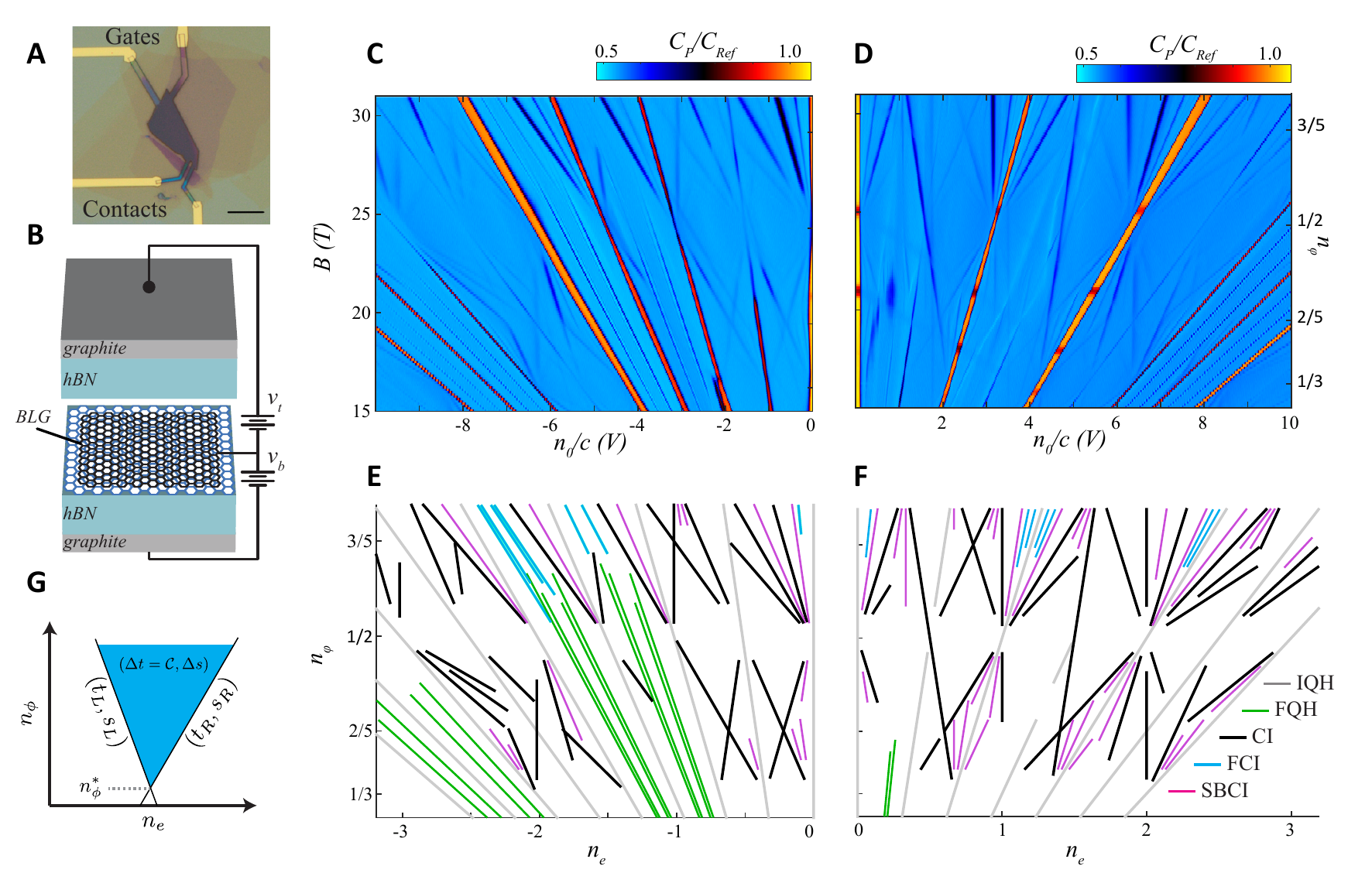} 
\caption{\textbf{Magnetocapacitance in a high-quality bilayer graphene moir\'e superlattice device.}
\textbf{(A)} Optical micrograph of the device. Scale bar is 10 $\mu$m. 
\textbf{(B)} Schematic of the device, with top and bottom graphite gates  at potential $v_t, v_b$. A moir\'e potential is induced by alignment of the graphene bilayer with one of the encapsulating hBN crystals.
\textbf{(C)} Penetration field capacitance ($C_{P}$)  as a function of density  $n_e\sim  n_0\equiv c(v_t+v_b)$ and magnetic field $B$ for $n_0<0$. T=300mK, and $C_{\textrm{Ref}}$ is a reference capacitance.
A large electric field $p_0/c = (v_t-v_b)$=16V is applied to force the valence electrons onto the top layer, which is in contact with the aligned hBN.
\textbf{(D)} $C_P$ for $n_0>0$ with $v_t-v_b$=-16 V at T=300 mK.
\textbf{(E-F)} Linear gap trajectories observed in (C-D) parameterized by $n_e=t\cdot n_\phi+s$.  $n_\phi$ and $n_e$ are the magnetic flux quanta and number of electrons per moir\'e unit cell, respectively. $n_\phi \equiv \frac{\sqrt{3} \lambda^2 B}{2 \Phi_0} = 1/2$ when B=24.3 T and $n_e = 1$ when $n_0/c = 3.1 V$.  Five trajectory classes are distinguished by color: Integer quantum Hall (gray, $s=0$, $t\in \mathbb{Z}$), fractional quantum Hall (green, $s=0$, $t$ fractional), Hofstadter Chern insulators (black, $s,t\in\mathbb{Z}$, $s\neq 0$), symmetry-broken Chern insulators (magenta, fractional $s$, $t\in\mathbb{Z}$) and fractional Chern insulators (cyan, fractional $s,t$).
\textbf{(G)} Schematic of a $(\Delta t, \Delta s)$ Chern band (see main text).
\label{fig1}}
	\end{center}
\end{figure*}

We measure the penetration field capacitance\cite{eisenstein_negative_1992} ($C_P$), which distinguishes between gapped, incompressible and ungapped, compressible states (see Supplementary Information).
Figs.~1C-D show $C_P$ measured as a function of $B$ and the electron density, $n \sim n_0\equiv c(v_t+v_b)$, where $v_t$ and $v_b$ are the applied top and bottom gate voltages and $c$ denotes the geometric capacitance to either of the two symmetric gates.
We use a perpendicular electric field, parameterized by $p_0/c=v_t - v_b$, to  localize the charge carriers onto the layer with a superlattice potential, e.g. adjacent to the aligned hBN flake.
High $C_P$ features, corresponding to gapped electronic states, are evident throughout the experimentally accessed parameter space (Fig.~1C-D), following linear trajectories in the $n - B$ plane. 
We estimate the area of the superlattice unit cell from zero-field capacitance data(see Supplementary Information), and define the electron density  $n_e = N_e / N_{\textrm{S}}$ and flux density $n_\Phi = N_\Phi / N_{\textrm{S}}$ per unit cell. Here $N_e$, $N_s$, and $N_\Phi$ are the number of electrons, superlattice  cells, and magnetic flux quanta in the sample, respectively.
The trajectories are parameterized by their inverse slope $t$ and $n$-intercept $s$ in the n-B plane,
\begin{equation}\label{eq:1}
N_e = t N_\Phi + s N_{\textrm{S}}, \quad \quad  n_e = t n_{\Phi} + s.
\end{equation}
The St\v{r}eda\cite{Streda} formula, $t=\frac{\partial n_e}{\partial n_\Phi} \big{|}_{N_S} =\frac{h}{e^2} \sigma_{xy}$, shows that the Hall conductance of a gapped phase is exactly $t$. 
The invariant $s = \frac{\partial N_e}{\partial N_S} \bigg{|}_{N_{\Phi_0}}$ encodes the amount of charge ``glued'' to the unit cell, i.e., the charge which is transported if the lattice is dragged adiabatically\cite{Macdonald}.
Non-zero $s$ indicates that strong lattice effects have decoupled the Hall conductance from the electron density.
Within band theory, the invariants of a gap arise from summing those of the occupied bands,
$(t, s) = \sum_{j \in \textrm{occ}} (\Delta t_j, \Delta s_j)$, and in particular the Hall conductance $t$ is the sum of the occupied band Chern indices, $\Delta t_j = \mathcal{C}_j$.

\begin{figure*}[ht!]
\begin{center}
\includegraphics[page=2,width=4.6in]{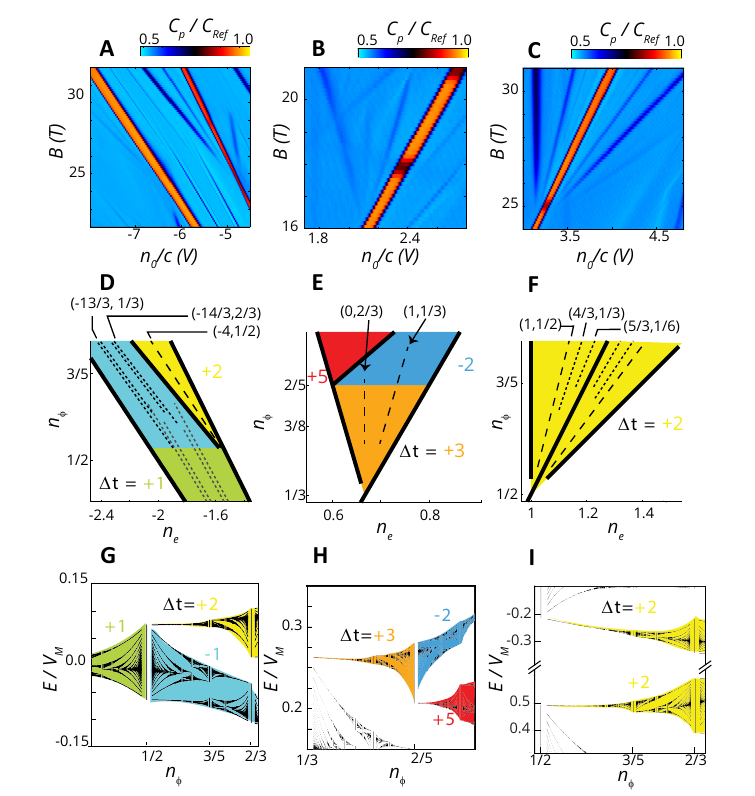} 
\caption{
\textbf{Interaction driven states at partial Chern-band filling.} 
\textbf{(A-C)} Details of Fig.~1C, D showing (A) FCI states in a $\Delta t$=-1 band,  (B) SBCI states in a $\Delta t$=3 band, and (C) FCI and SBCI states in $\Delta t$=2 bands.
\textbf{(D)} Schematic of (A).  FCI states (dotted lines) with $(t,s) = (-13/3,1/3)$ and $(-14/3,2/3)$ occur at fractional filling of a $\Delta t = -1$ band (light blue). 
\textbf{(E)} Schematic of (B).  SBCI states (dashed lines)  at $(t,s) = (0,2/3)$ and $(1,1/3)$ occur at 1/3 and 2/3 fractional filling of a $\Delta t$=3 band (orange). 
\textbf{(F)} Schematic of (C).  Both FCI and SBCI states (dotted and dashed lines) occur in the $\Delta t$=2 bands
\textbf{(G)} Calculated Hofstadter energy spectrum in the regime of (A), matching the observation that the LL splits into $C=-1, 2$ bands.
\textbf{(H)} Calculated Hofstadter spectrum in the regime of (B), matching the observed splitting of a $C=$3 band into  $C=5,-2$ bands. 
\textbf{(I)} Calculated Hofstadter spectrum in the regime of (C). The IQH gap at $\nu = 2$ separates the two single-particle bands and is much larger than $V_M$. \label{fig2}}
\end{center}
\end{figure*}

We observe five classes of gap trajectories based on the properties of $t$ and $s$, each of which correspond to a distinct class of insulating state (Fig.~1E-F).  Free-fermion states must have integer $t$ and $s$: trajectories with $s=0$ correspond to IQH effects between LLs, while trajectories with $s\neq0$ indicate the formation of the non-LL Chern bands of the Hofstadter butterfly\cite{Dean13Nat,Ponomarenko13Nat,Hunt13Sci}.
Fractional $t$ or $s$ are beyond the single particle picture and thus indicate interaction-driven phases. The conventional FQH states follow trajectories with fractional $t$ and $s=0$. Gaps with integer $t$ and fractional $s$ (previously observed in monolayer graphene \cite{Wang15Sci}) must be either topologically ordered \emph{or} have interaction-driven spontaneous symmetry breaking of the superlattice symmetry. The theoretical analysis below suggests the latter case is most likely, so we refer to this class as symmetry-broken Chern insulators (SBCIs).  Finally, there are gaps with fractional $t$ and fractional $s$, which are the previously unreported class of topologically-ordered FCI phases.

To better understand states with fractional $t$ or $s$, we first identify the single-particle Chern bands in our experimental data by identifying all integer-$t$, integer-$s$ gaps. We focus on adjacent pairs of integer gaps, $(t_L, s_L)$ and $(t_R, s_R)$, which bound a finite range of $n_e$ in which no other single-particle gaps appear (Fig.~1G). Adding charge to the left gap corresponds to filling a Chern band with invariants  $(\Delta t,  \Delta s) = (t_R - t_L, s_R - s_L)$.
From this criterion we find a variety of Chern bands with $\Delta t =\pm 1, \pm 2, \pm 3$ and $\pm 5$ in the experimental data(see Supplementary Information), each of which appear as  a ``triangle'' between adjacent single particle gaps. These Chern bands are observed to obey certain rules expected from the Hofstadter problem: for example, $\Delta t$ and $\Delta s$ are always coprime, and Chern-$\Delta t$ bands always emanate from a flux $n^\ast_\phi = p / \Delta t$.

Interaction-driven phases occur at fractional filling $\nu_\textrm{C}$ of a Chern band, following trajectories $(t_{\nu_\textrm{C}}, s_{\nu_\textrm{C}}) = (t_L, s_L) + \nu_\textrm{C}  (\Delta t, \Delta s)$.  
The Chern numbers of the bands in which some of the observed interaction-driven phases appear (Figs.~2A-C) are denoted schematically in Figs.~2D-F.

By combining a phenomenological description of the moir\'e potential with knowledge of orbital symmetry breaking in bilayer graphene\cite{Hunt16}, we are able to construct a single particle model which closely matches the majority of the experimentally observed single-particle Chern bands(see Supplementary Information).
The calculated energy spectra of the bands relevant to Figs.~2A-C are shown in Fig.~2 G-I. As is clear from the band structure, stable phases at fractional $\nu_\textrm{C}$ are not expected within the single particle picture: instead, the encompassing Chern band splits indefinitely into finer Chern bands at lower levels of the fractal butterfly that depend sensitively on $n_\phi$.

The three columns of Fig.~2 represent instances of three general classes of  fractional $\nu_c$ states observed in our experiment.  
Fig. 2A shows two gapped states within a $\Delta t=-1$ band at $\nu_\textrm{C}=\frac{1}{3}$ and $\frac{2}{3}$. These gaps extend from $n_\phi \approx 0.55$ to at least $n_\phi=0.8$(see Supplementary Information).  Both are characterized by fractional $t$ and $s$, and we identify them as FCI states. As with FQH states, the fractionally quantized Hall conductance implies that the system has a charge $e/3$ excitation\cite{laughlin_anomalous_1983}. The  fractional $s$ values of these states, being multiples of this fractional charge, do not require broken superlattice symmetry. An analogy between Laughlin states in a conventional LL, shown in Fig. 3A, and FCI in a $\Delta t=-1$ band is suggested by the observation of apparent FCI hierarchy states at $\nu_C=2/5,3/5$ (Fig.~3B).  

\begin{figure}
\begin{center}
\includegraphics[page=3,width=2.3 in]{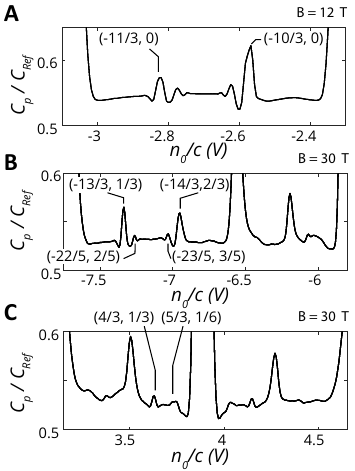} 
\caption{
\textbf{Line cuts of $C_P$ comparing FCI and FQH states.}
\textbf{(A)} Line cut averaged over $p_0/c \sim 1.0 - 4.0 V$ at $B=12$ T, showing FQH states in a conventional LL.  At low fields, the effective moir\'e potential is weak, and FQH states are observed at $\nu_C=1/3, 2/3$ as well as $2/5, 3/5$ of the $\Delta t=+1$ LL. 
\textbf{(B)} Line cut averaged over $p_0/c \sim 4.0-14.0 V$ at $B=30$ T, showing FCI in the same $\Delta t=-1$ band as in Fig. 2A. Weaker features appear at 2/5 and 3/5 filling of the band, similar to the Laughlin sequence in (A). 
\textbf{(C)} Line cut averaged over $p_0/c \sim -14.0$ to $-9.0 V$ at $B=30$ T, showing FCI  in the same $\Delta t=2$ band as shown in  Fig.~2C. The relative strength of the (4/3,1/3) state compared to the (5/3,1/6) state is consistent with the former preserving the lattice symmetry.  
\label{fig_linecuts}}
\end{center}
\end{figure}

Fig.~2B shows gapped states in a $\Delta t=+3$ band at $\nu_{\textrm{C}}=1/3, 2/3$, while Fig.~2C shows gapped states in two $\Delta t=+2$ bands at $\nu_{\textrm{C}}=1/2$.  Filling a Chern-$\Delta t$ band to a multiple of $\nu_{\textrm{C}} = \frac{1}{|\Delta t|}$ corresponds to integer $t$ but fractional $s$.  While we cannot exclude exotic fractionalized states at these fillings these states are unlikely to admit a simple interpretation as FCIs. Absent fractional excitations, a gapped state with fractional $s = \frac{x}{y}$ implies broken superlattice symmetry: the unit cell of such a phase must contain an integral number of electrons, and the smallest such cell contains $y$ superlattice sites. Theoretically, such symmetry breaking is expected to arise spontaneously due to electronic interactions, in a lattice analog of quantum Hall ferromagnetism\cite{Kumar}.
A $\Delta t$ Chern band is similar to a $\Delta t$-component LL, but in contrast to an internal spin, translation acts by cyclically permuting the components\cite{Barkeshli,Kumar,WuBernevigRegnault}.
Spontaneous polarization into one of these components thus leads to a $t$-fold increase of the unit cell\cite{Kumar}. 
The observation of SBCIs is thus analogous to the observation of strong odd-integer IQHEs which break spin-rotational invariance.  
Some of the ``fractional fractal'' features recently described in monolayer graphene appear to be consistent with this explanation\cite{Wang15Sci}.

Finally, we also observe fractional-$t$ states within a $\Delta t=+2$ band (Fig. 3C), for example at $\nu_C=2/3$ ($s=4/3$ and $t=1/3$) and $\nu_C=5/6$ ($s=5/3$ and $t=1/6$). FCIs in Chern-$\Delta t\neq1$ bands can either preserve or break the underlying  lattice symmetry.  Symmetry preserving FCIs are expected\cite{sterdyniak_series_2013, kolread, MollerCooper} at fillings $\nu_\textrm{C} = \frac{m}{2 l m \Delta t + 1}$ for integers $l, m$, consistent with the stronger $\nu_C=2/3$ states ($l = 1, m = -1$). At $\nu_C=5/6$, in contrast, the weak state observed is not consistent with this sequence. For this state, $t=5/3$ implies a likely fundamental charge of e/3, while $s=1/6$. By analogy to SBCIs, this implies that one half of the fundamental charge is pinned to each moir\'e unit cell, suggesting the unit-cell is doubled in this ``SB-FCI'' state. This scenario is again closely analogous to the physics of a spin degenerate LL; at 1/6 filling of a spin degenerate LL the system spontaneously polarizes into a $t = \frac{1}{3}$ Laughlin state, while at 1/3 filling the system can form a spin-singlet, $t = \frac{2}{3}$ state.

\begin{figure}[h!]
\begin{center}
\includegraphics[page=4,width=2.3 in]{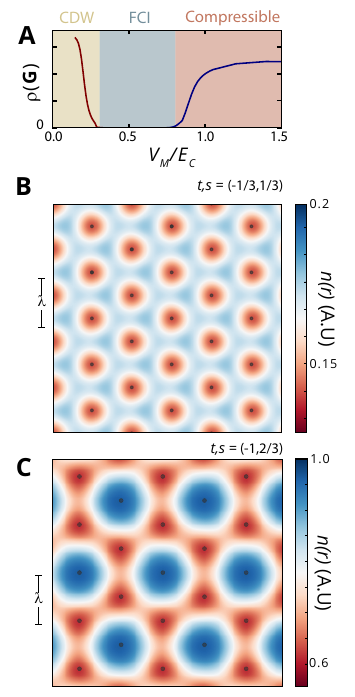} 
\caption{
\textbf{iDMRG calculations showing the stability of FCI and SBCI states}
\textbf{(A)} Calculated iDMRG phase diagram at $\nu_\textrm{C} = $ 1/3 filling of the $\Delta t$ = -1 band shown in Fig.~2A,D,G ($n_\phi$ = 2/3).  $|V_M|$ is the moir\'e potential amplitude, $E_C$ is the Coulomb energy, and $\rho(\mathbf{G})$ is the charge density at Bragg vector $\mathbf{G}$.
The FCI competes with two other phases: a charge density wave (CDW) at low $|V_M|$, and a compressible phase at high $|V_M|$.
The competing phases are diagnosed by symmetry breaking density waves at wavevector $\mathbf{G}$ = $\mathbf{G_0}/3$ (red) and $\mathbf{G}$ = $\mathbf{G_0}/2$ (blue), where $G_0$ is a reciprocal vector of the moir\'e(see Supplementary Information).
\textbf{(B)} Calculated real-space electron density $n(r)$ of the FCI found in (A).  $n(r)$ preserves the symmetry of the moir\'e potential, whose periodicity is indicated by the gray circles. Here $V_M/E_C = 0.7$.
\textbf{(C)} Calculated real-space electron density $n(r)$ at $\nu_C = \frac{2}{3}$ filling of the $\Delta t$ = 3 band shown in Fig.~2B,E,H ($n_{\phi} = 3/8$).
The result is consistent with an SBCI phase; $(t,s) = (-1,2/3)$, while $n(r)$ spontaneously triples the unit cell of the underlying moir\'e potential, indicated by gray circles.  Here  $V_M/E_C = 0.6$ and $\Theta_M = \pi/8$.  \label{fig4}}
\end{center}
\end{figure}

To assess the plausibility of FCI and SBCI ground states, we use the infinite density matrix renormalization group (iDMRG) to numerically compute the many body ground state within a minimal model of the BLG\cite{ZaletelMongPollmannRezayi}. We first consider Coulomb interactions and a triangular moir\'e potential of amplitude $V_M$ projected into a BLG $N=0$ LL\cite{chen_zero-energy_2016}, matching the parameter regime in Fig.~2A (see Supplementary Information).  We focus on $n_\phi = \frac{2}{3}$ at a density corresponding to $\nu_{\textrm{C}} = \frac{1}{3}$ filling of the $\Delta t=-1$ band.

If interactions are too weak compared to the periodic potential (as parameterized by $V_M/E_C$, where $E_C=e^2/(\epsilon\ell_B)$ is the Coulomb energy, $\ell_B = \sqrt{\frac{\hbar}{eB}}$ is the magnetic length, and $\epsilon$ the dielectric constant) the ground state at $n_\phi = \frac{2}{3}$ is gapless, corresponding to a partially filled Chern band. If the interactions are too strong, the system forms a Wigner crystal which is pinned by the moir\'e potential. In the intermediate regime, however, 
the numerical ground state of this model has a fractional $t$ and $s$ which match the experiment, and hence is an FCI, with entanglement signatures that indicate a Laughlin-type topological order.
The FCI is stable across an range of $V_M/E_C$ (Fig.~4A) corresponding to $|V_M|\approx 14-38$~meV, consistent with recent\cite{LeeScience2016} experiments that suggest $|V_M| \sim 25$ meV. Fig.~4B shows that the real-space density of the FCI is strongly modulated by the potential, but preserves all the symmetries of the superlattice.

We next conduct iDMRG calculations to assess the plausibility of the SBCI hypothesis.
We focus on the well developed Chern-$3$ band of Fig.~2B,E,H.
As a minimal model, we project the moir\'e and Coulomb interactions into the $N=1$ LL of the BLG, fixing $V_M = 21$ meV and $E_C(B=17T) = 35$meV, and take $n_\phi = \frac{3}{8}$.

At $\nu_\textrm{C} = \frac{1}{3}$ filling, the electron density indeed exhibits a modulation which spontanesouly triples the superlattice unit cell (Fig.~4C).
A similar tripling is observed at $\nu_\textrm{C} = \frac{2}{3}$.
These are not merely density waves, however, as they have finite $(t, s)$ invariants, in agreement with experiment.

We note that the SBCI states are distinct from a second class of integer-$t$, fractional-$s$ features, the moir\'e-pinned Wigner crystals\cite{Wang15Sci, DaSilva}.
In the latter case, starting from a LL-gap at $t, s = t_L, 0$, additional electrons form a Wigner crystal pinned by the moir\'e; the added electrons are electrically inert, leading to a state at $t, s = t_L, \frac{x}{y}$ which can't be ascribed to fixed $\nu_\textrm{C}$ of an encompassing band. These states are thus analogous to reentrant IQH effects, with the moir\'e playing the role of disorder.
In contrast, while the electrons added to the SBCI spontaneously increase the unit cell, they also contribute an integer Hall conductance, which together correspond to some $\nu_\textrm{C}$.

In summary, we find that instead of a self-repeating fractal structure, interactions mix Hofstadter-band wavefunctions to form stable, interaction-driven states at fractional filling of a Chern band. Among these are both symmetry-broken Chern insulators and topologically-ordered fractional Chern insulators, the latter of which constitute a lattice analog of the FQH effect. 
FCIs provide new avenues to experimental control through lattice engineering, particularly in higher Chern number bands where candidate two component states, such as that observed  at $\nu_C=2/3$ of the $\Delta t=2$ band of Fig. 3C, may host nonabelian defects at engineered lattice dislocations\cite{Barkeshli}. A pressing experimental question is thus whether FCI states can be realized in microscopically engineered superlattices.

\tocless\section{Acknowledgments}
The authors acknowledge discussions with Maissam Barkeshli, Andrei Bernevig, Cory Dean, and Roger Mong and experimental assistance from Jan Jaroszynski and Matthew Yankowitz. The numerical simulations were performed  on computational resources supported by the Princeton Institute for Computational Science and Engineering using iDMRG code developed with Roger Mong and the TenPy Collaboration.
EMS acknowledges the support of the Elings Fellowship. K.W. and T.T. acknowledge support from the Elemental Strategy Initiative
conducted by the MEXT, Japan and JSPS KAKENHI Grant Number JP15K21722.
Measurements were performed at the National High Magnetic Field Laboratory, which is supported by National Science Foundation Cooperative Agreement No. DMR-1157490 and the State of Florida. 
The work at UCSB was funded by ARO under proposal 69188PHH.   
AFY acknowledges the support of the David and Lucile Packard Foundation.

\clearpage

\renewcommand{\emph}{\textit}
\renewcommand{\thefigure}{S\arabic{figure}}
\renewcommand{\thetable}{S\arabic{table}}
\renewcommand{\theequation}{S\arabic{equation}}

\begin{widetext}
\vspace*{0.5cm}
\textbf{\large Supplementary Online Material: Observation of fractional Chern insulators in a van der Waals heterostructure}
\vspace*{0.5cm}
\end{widetext}

\tableofcontents

\section{Experimental methods}
The device fabrication and measurement techniques presented in this manuscript are identical to those presented in Ref.~\cite{zibrov_tunable_2017}, and the device we study here is Sample A in that manuscript. A more comprehensive discussion of fabrication and experimental methods is found in the supplementary material of that paper\cite{zibrov_tunable_2017}.

We assembled the heterostructure using a dry transfer method which utilizes the van der Waals force to fabricate layered structures consisting of hBN, graphite, and graphene. We contacted the bilayer graphene directly with a thin graphite contact, which in turn was edge contacted with Cr/Pd/Au metallic leads\cite{wang2013one}. The top and bottom gates are also thin graphite, which results in devices with significantly less disorder than similar heterostructures with metal gates made using standard deposition techniques\cite{zibrov_tunable_2017}.

We performed magnetocapacitance measurements to identify bulk gapped states as described in Ref. \onlinecite{zibrov_tunable_2017} and references therein. In this work we measure the penetration field capacitance $C_P$ and the symmetric capacitance $C_S$, both of which primarily access whether the bulk of the device is gapped or not. $C_P$ is the capacitance between the top and bottom gate, and it is suppressed when the bilayer can screen electric fields (i.e. when the bilayer is compressible and conducting). Gapped states, therefore, appear as peaks of enhanced $C_P$. $C_S$ is the sum of the capacitances of the bilayer to the top and bottom gates, is suppressed when the bilayer is more insulating/incompressible, and therefore gaps appear as dips in $C_S$. We have chosen the color scale for both $C_S$ and $C_P$ such that gaps appear as warmer colors, despite the sign difference of the gapped features.

Due to a small asymmetry between the top and bottom hBN thicknesses, we observed a corresponding asymmetry between top and bottom gate capacitances $\delta \equiv (c_t - c_b) / (c_t + c_b) = 0.018$ which was taken into account when applying $n_0/c$ and $p_0/c$ to the device. The full expressions including this asymmetry are $n_0/c = (1-\delta)v_t + (1+\delta)v_b$ and $p_0/c = (1-\delta)v_t - (1+\delta)v_b$.

The data presented here was taken at relatively high frequencies (between 60 and 100 kHz), where an out of phase dissipative signal is present in many of the gapped states we observe.  This arises because the measurement time is not sufficient to fully charge the sample. In this regime, measured capacitance is a convolution of both conductivity and compressibility\cite{goodall_capacitance_1985}; however because both low conductivity and low compressibility are hallmarks of gapped states, this does not affect the interpretation of high $C_P$ or low $C_S$ as indicative of a gapped state.

We performed the magnetocapacitance measurements at the National High Magnetic Field Lab in He-3 refrigerators at their base temperature of T $\sim 300$ mK. In both measurements, we ramped the field continuously while performing the measurements and were unable to concurrently record the magnetic field. There are systematic errors in the reported field up to $\sim$ 0.5 T between different data sets due to errors in timing between data acquisition and the field sweep.

\begin{figure*}
\begin{center}
\includegraphics[width=4.5in]{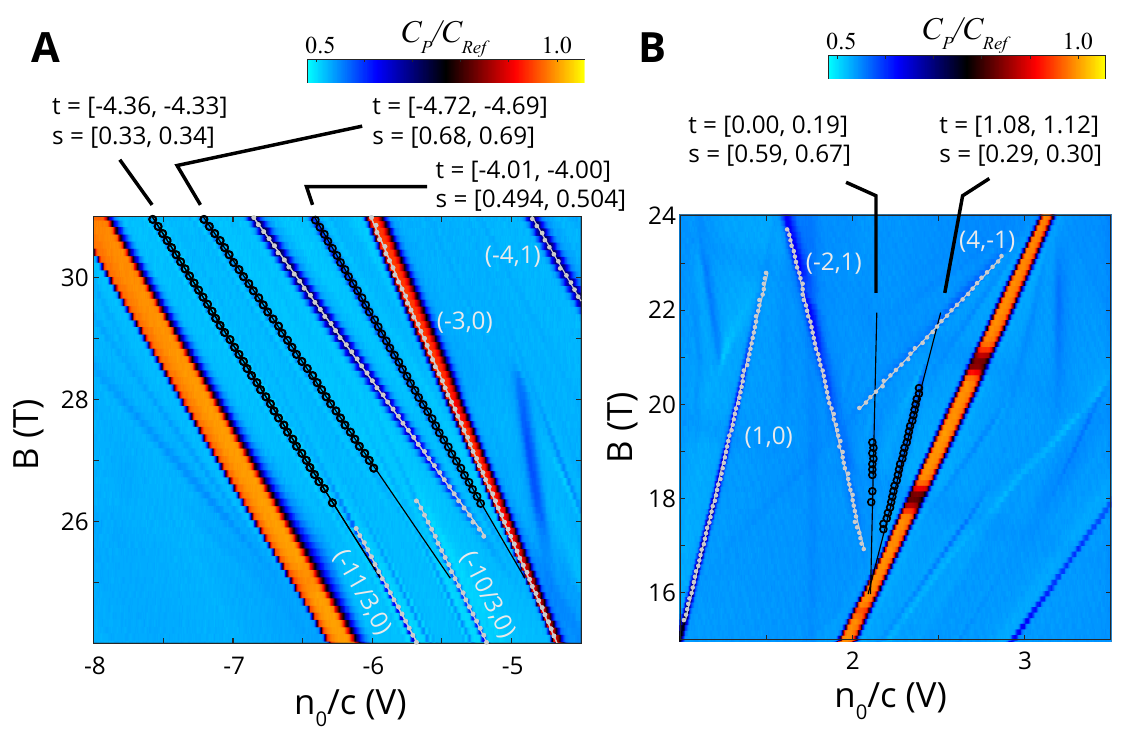} 
\caption{
\textbf{Fits to $t,s$ for FCI and SBCI states.}
\textbf{(A)} Peaks corresponding to the FCI states presented in Fig.~2A,D (black circles) were fit to obtain their slope and intercept in the $n_0/c-B$ plane (black lines). The fitted slopes of nearby CI and FQH features (gray lines) were used to convert from the fitted parameters to a fitted $t,s$, which match the expected values of (-13/3,1/3) and (-14/3,2/3) to within 5\%. Numbers in brackets are the 95\% confidence intervals obtained by linear regression. An SBCI in a $\delta t=2$ band also matches its predicted value of $(t,s)$ =(-4, 0.5).  
\textbf{(B)} Similar analysis performed for SBCI states in Fig.~2B,E. $t,s$ values for these SBCI match within 10 \% of their expected values of (0,2/3) and (1, 1/3).  
\label{fig:fits}}
\end{center}
\end{figure*}

We identify $t,s$ of linear gap trajectories in the main text  by visually comparing slopes to known features such as IQH gaps and identifying fields at which multiple features intersect. To more robustly confirm the finding of fractional $t,s$ states, we used a peak finding algorithm to identify peaks in each horizontal line scan of $C_P$ (see Fig.~\ref{fig:fits}), manually grouped the peaks belonging to a single trajectory and then fit their slope and intercept in the $n_0$-$B$ plane. Quantum capacitance prevents a direct conversion from the fitted slope and intercept to quantitative $t,s$ for the full range of measured voltages and fields. Therefore we used the fitted slopes and intercepts of nearby nearby CI and FQH features to obtain to fix the local conversion to $t,s$. These local conversions also give a quantitative check on the conversions from $B$ to $n_\phi$ and $n_0/c$ to $n_e$ used in the main text. For Fig.~\ref{fig:fits}A, we find that $B_{\Phi_0}$ = 48.6 T and $n_0/c = $ 3.08 V at $n_e$ = 1 and for Fig.~\ref{fig:fits}B we find $B_{\Phi_0}$ = 48.3 T and $n_0/c = $ 3.10 V at $n_e$ = 1. Both of these conversions are consistent with the values used in Figs.~1E,F.

\section{Estimating the moir\'e periodicity}

 \begin{figure}
 \begin{center}
\includegraphics[width=3in]{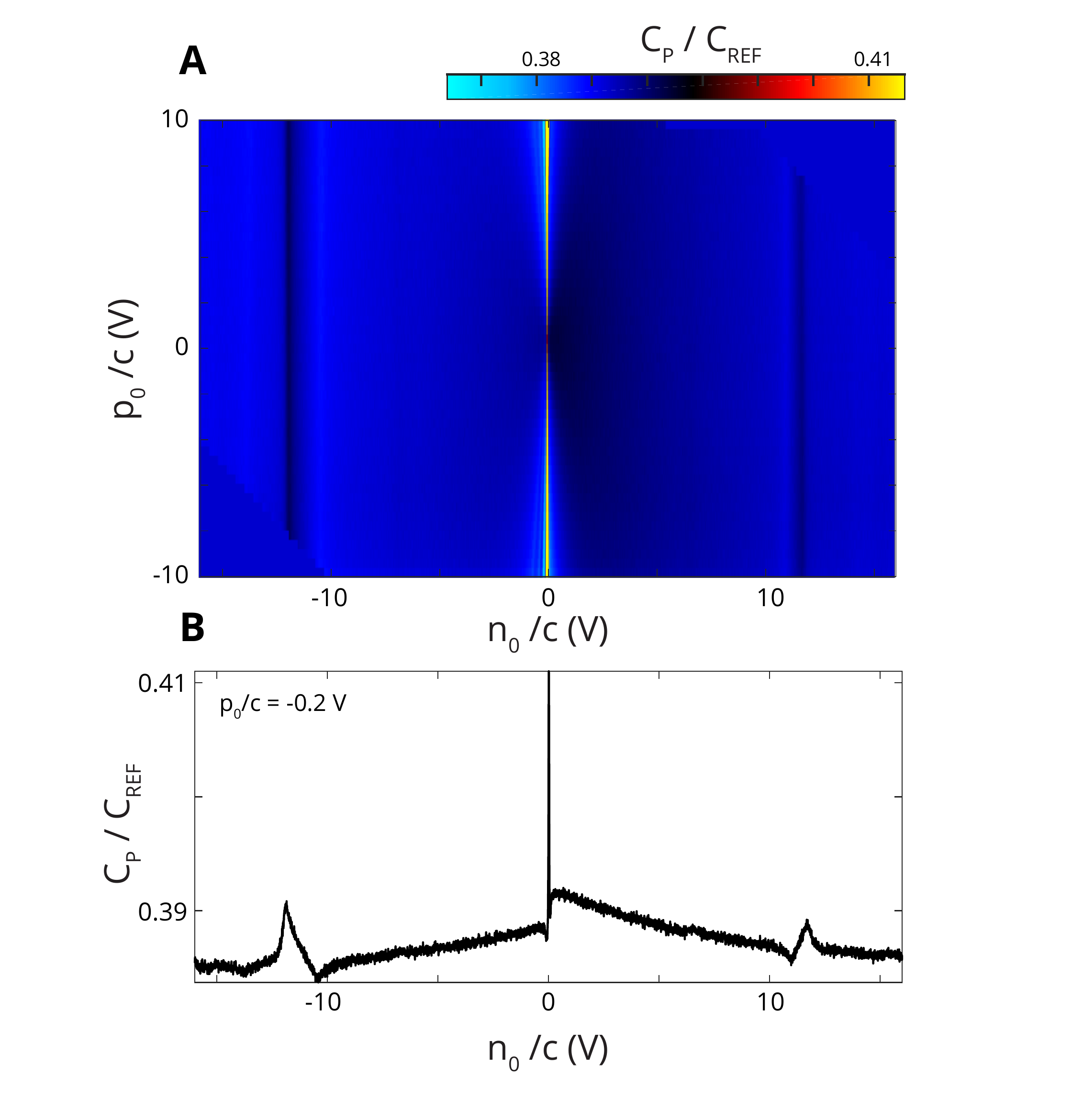} 
\caption{\textbf{Zero magnetic field $C_P$ showing satellite Dirac points.}
\textbf{(A)} $C_P$ taken as a function of the nominal electron density ($n_0/c$) and polarizing electric field ($p_0/c$). This measurement was taken at the nominal base temperature of our dilution refrigerator ($T < 50$ mK) at zero applied magnetic field.  We find additional peaks in $C_P$ at $n_0/c \approx$ 11.8 V. The top right and bottom left corners are masked off (data was not taken in those regions)  to protect the gates from leakage. \textbf{(B)} Horizontal line cut of (A) taken at fixed $p_0/c=-0.2 V$. 
\label{fig:satpeaks}}
 \end{center}
 \end{figure}

The encapsulated nature of our device does not allow direct scanned probes of the moir\'e pattern, so we must rely on electronic signatures of the superlattice. First, we estimate the periodicity from zero field features in the density of states and the geometry of our device. We observe satellite peaks in $C_P$ at approximately $|n_0^{sat}/c| =$ 11.8 $V$, which do not vary strongly with $p_0$ (Fig.~\ref{fig:satpeaks}). These peaks are  a direct consequence of the moir\'e periodicity and occur at $n_e$ = $\pm 4$ in bilayer graphene, e.g. when there is one of electron of each spin-valley flavor per moir\'e unit cell \cite{Dean13Nat,Hunt13Sci,Ponomarenko13Nat}.
For a triangular lattice
\begin{equation}
e g_s g_v \frac{2}{\sqrt{3}\lambda^2} = \frac{\epsilon n_0^{sat}/c}{d}
\end{equation}
where $g_s=2$ and $g_v=2$ are the spin and valley degeneracies, $\lambda$ is the moir\'e wavelength, $\epsilon$ $= (3\pm0.15) \epsilon_0$ is the dielectric constant for hBN \cite{Hunt16}, $d = 45$nm is the average thicknesses of the two hBN flakes, and $n_0^{sat} /c $ is the value of $v_t + v_b$ at which satellite peaks appear. We estimate, therefore, that $\lambda = 10.3 \pm 0.3$ nm and predict $n_\phi = 1$ should occur at $B = \Phi_0 / (\sqrt{3} \lambda^2 / 2) = 42.5-47.8$ T.

A more accurate method for determining the moire potential is by noting the crossing of many trajectories around $B =$ 24.3$\pm$.2 T, and associating this field with as $n_\Phi =$ 1/2.  This implies a moir\'e periodicity of $\lambda = 9.92\pm.03$nm, consistent with the zero-field assessment but considerably more precise.  Unlike the zero field assessment, the latter estimate is less susceptible to quantum capacitance corrections to the realized density.  Note that for analysis of the observed trajectories in $C_P$ described in the main text, $t$ (the inverse slope) is unaffected by the choice of $\lambda$, as both $n_e$ and $n_\Phi$ go as $\lambda^{-2}$.

 \begin{figure*}[ht!]
 \begin{center}
\includegraphics[width=6.5in]{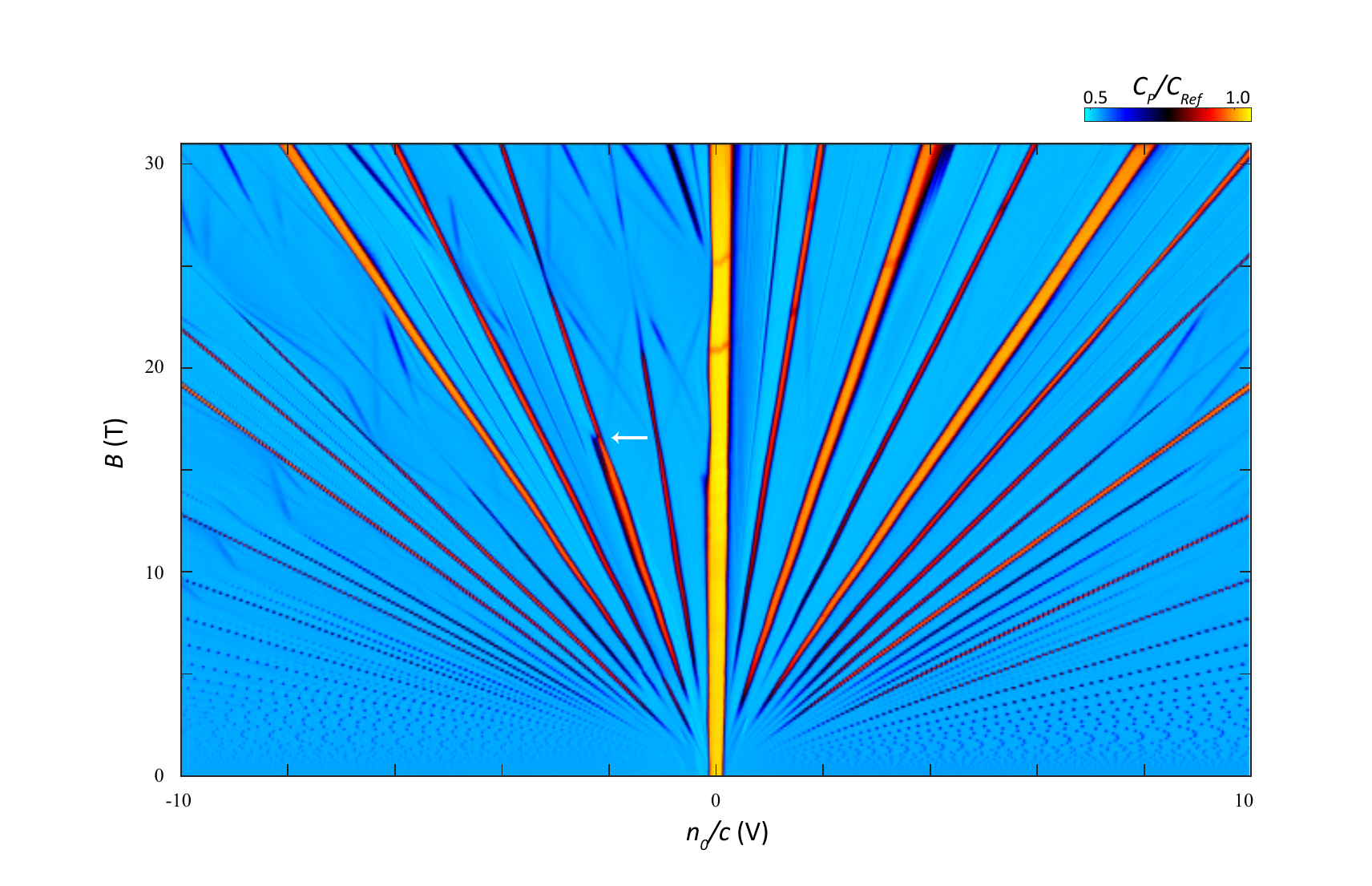} 
\caption{\textbf{Full Landau fan at $p_0/c$ = +16 V.}
The effect of the moir\'e between the top hBN and bilayer graphene is largely suppressed for states localized on the bottom layer (e.g. when $n_0>$ 0, $p_0/c$ = 16 V). In this regime, Landau levels in the ZLL exhibit the conventional fractional quantum Hall effect up to 31 T and do not show any Hofstadter (non-zero $s$) features. Hofstadter features are, however, observed in higher Landau levels in this regime. 
Additionally, we observe a feature $B \sim$ 16 T and $\nu = -2$ (white arrow) which we attribute to a spin transition between a spin-unpolarized $\nu = -2$ at low fields to a spin-polarized $\nu = -2$ at high fields, as described in the supplementary text.  Above this field, the ZLL fills two N=0 orbital states first (rather than an N=0 followed by an N=1). This feature, not previously reported, can be attributed to the very large electric fields used in this experiment as compared with previous work\cite{Hunt16}.
\label{fig:full_fan}}
 \end{center}
 \end{figure*}

\section{A minimal model for BLG with a single-layer moir\'e potential}

The interplay between the moir\'e potential and the complex high-field physics of bilayer graphene is non-trivial.
There are a large number  of degrees of freedom (spin, valley, and LL-level index)  and competing energy scales (the cyclotron energy $\hbar \omega_c$, the Coulomb scale $E_C = e^2 / \epsilon \ell_B$,  the potential bias across the bilayer $u$, the Zeeman energy $E_Z$, the amplitude of the moire potential $V_M$, and various small interaction anisotropies).
In particular, interactions are essential, and even integer gaps cannot be understood based on a single particle model\cite{Hunt16}.
While a complete understanding at the microscopic level is not required to demonstrate fractional filling of Chern bands, which follows purely from the observation of quantized fractional $s$ and $t$, in this section we motivate an approximate model for the system which is the starting point for our DMRG simulations. 
A number of features of our data can be accounted for in this model, including the dominant single-particle CI features.

\subsection{The ZLL in the absence of a moir\'e potential}   

The LLs of graphene are labeled by the electron spin ($\sigma = \pm \frac{1}{2}$), the graphene valley index ($\xi =  + / -$), and the integer LL index ($N \in \mathbb{Z}$).
The spin and valley combine to form an approximately SU(4)-symmetric ``isospin,'' and so the order in which the levels fill depends on various competing anisotropies.
Of particular interest are the eight components of the zeroth Landau level (ZLL), which includes both $N=0, 1$ and fills for $-4 < \nu <  4$,  the regime of our experiment.
A detailed experimental and theoretical account of the ZLL in the absence of a moir\'e was provided in Ref.~\onlinecite{Hunt16}, to which we refer the interested reader.
Here, we summarize those results at a  qualitative level in order to argue the following:  
\begin{itemize}
\item[(1)] it is a reasonable starting point to project the problem into the eight degrees of freedom of the ZLL
\item[(2)] because of the large interlayer potential difference $u$ applied in the current experiment, it is further justified to restrict to the four ZLL levels in valley $+$, i.e., $\ket{\xi N \sigma} \in \{\ket{+ 0 \uparrow}, \ket{+ 0 \downarrow}, \ket{+ 1 \uparrow}, \ket{+ 1 \downarrow} \}$
\item[(3)] these levels fill in a different order depending on whether $u < 0$ or $u > 0$, leading to different Chern bands and fractional states in these two regimes.
\end{itemize}
    
\paragraph{ZLL projection.}
To a good approximation, the cyclotron energies of BLG scale  as $E^{(\omega_c)}_N \approx \hbar \omega_c \sqrt{N (N-1)}$. 
The $N=0, 1$ levels are near degenerate, so together with spin and valley combine to give the eight components of the ``zeroth Landau level'' (ZLL).
In our experiment, the cyclotron splitting is $\hbar \omega_c \sim 45 - 120$meV across the range $B \sim 17 - 44$T, the Coulomb interactions are at scale $E_C \sim 35 - 57$meV. Earlier electron focusing experiments\cite{LeeScience2016} on encapsulated monolayer graphene estimated the moir\'e potential magnitude as $|V_M| \sim 10 - 20$meV.
Given the hierarchy of scales $E_C, V_M < \hbar \omega_c$, it is reasonable to project the problem into the eight components of the ZLL. Note that even with the large bias $u$ (controlled by the experimental parameter $p_0$\cite{Hunt16}) applied in our experiment, we do not observed crossings between the ZLL and higher LLs.

Focusing on the ZLL, which fills from $-4 < \nu < 4$, the single particle energies take the form \cite{Hunt16}
\begin{align}
E^{(1)}_{\textrm{ZLL}} &= N \Delta_{10} - \sigma E^{(Z)}  - \xi \tfrac{u}{2} \alpha_N \\
&\approx  B  \left[ \frac{ \textrm{meV}}{\textrm{T}} \right] \left[ N (0.3  + \xi \frac{u}{2} 0.013)  - \sigma \, 0.116   - \xi \frac{u}{2} \right] 
\end{align}
Here $1 = \alpha_{N=0} > \alpha_{N=1} > 0$ and $\Delta_{10} > 0$ are $B$-dependent factors which can be computed numerically from the band structure of bilayer graphene, $u$ is the potential difference between the two layers due to a perpendicular electric field (in meV),  and $E^{(Z)}$ is the Zeeman splitting.

\paragraph{Restriction to valley $+$.} The effect of the bias $u$ depends directly on the valley $\xi = \pm$; this is because the ZLL wavefunctions have the  property that valley $+$ is largely localized on the top layer, while valley $-$ is largely localized on the bottom layer.  Within the ZLL, then,  valley $\approx$ layer and the bias $u$ splits the valley degeneracy.
In our experiment, the top and bottom gates are approximately 100nm apart and at a voltage difference of $\pm 16 V$. The layer separation of BLG is 0.335nm, so we expect a large bias $\frac{u}{2} \sim \pm 16 \frac{0.335}{100}$eV $ = \pm 27$meV across the bilayer, though the precise value of $u$ is modified somewhat due to the relative dielectric constant of the BLG and hBN. Regardless,  $|u|$ is large enough to ensure that for $u > 0$, valley $\xi = +$ fills before valley $\xi = -$, while for $u < 0$ the reverse occurs\cite{Hunt16}.
Since the moir\'e potential couples dominantly to the top layer (a consequence of the near perfect crystallographic alignment of the BLG with the top hBN, but misalignment with the bottom hBN), we  expect the Hofstadter features to appear most strongly when valley $+$ is filling. 
This is confirmed by the Landau fan at $u>0$, shown in Fig.~\ref{fig:full_fan}. For $n_0 < 0$, the top layer ($+$) is filling and we see very strong Hofstadter features, while for $n_0 > 0$, the bottom layer  ($-$) is filling and the Hofstadter features are absent or weak. For $u < 0$ (not shown), the opposite is observed.
For this reason, in the main text we present data for $n_0 < 0, u > 0$, and $n_0 > 0, u < 0$, in order to focus on the electrons affected by the moir\'e. We thus restrict our attention to the four components $+N\sigma$ of the ZLL.

\paragraph{$u$-dependence of the filling order.} For valley $+$, the splitting between the $N=0, 1$ orbitals is 
\begin{align}
\epsilon_{10} \sim B  \left[ \frac{ \textrm{meV}}{\textrm{T}} \right]   ( \tfrac{u}{2} 0.013 + 0.3 )  
\end{align}
Comparing with the small Zeeman energy, at the non-interacting level (for moderate $u$) we expect the levels $\ket{\sigma, N} $ to fill in the order  $\ket{\uparrow, 0}, \ket{\downarrow, 0}, \ket{\uparrow, 1}, \ket{\downarrow, 1}$.
However, Coulomb interactions rearrange this order, because filling two orbitals of the same spin, e.g.  $\ket{\uparrow, 0},  \ket{\uparrow, 1}$, has much more favorable Coulomb energy than filling two orbitals of opposite spin, e.g., $\ket{\uparrow, 0}, \ket{\downarrow, 0}$. Having filled $\ket{\uparrow, 0}$, this effectively reduces the energy of the $\ket{\uparrow, 1}$ level by an amount
$\xi_{10} \propto E_C \propto \sqrt{B}$  (at the level of Hartree-Fock, this is the difference in ``exchange energy'').
If $\epsilon_{10} - \xi_{10} < 0$, the orbitals will instead fill in order $\ket{\uparrow, 0}, \ket{\uparrow, 1}, \ket{\downarrow, 0}, \ket{\downarrow, 1}$, an effect which was confirmed experimentally in Ref.~\onlinecite{Hunt16}.
However, because $\epsilon_{10} \propto B$ while $\xi_{10} \propto \sqrt{B}$, there is a critical $B$ where $\epsilon_{10}$ wins out and the ordering should revert to that expected from the non-interacting picture. 
For $u \gg 0$ (e.g., region $n_0 < 0$), $\epsilon_{10}$ is large and this transition should occur at moderate $B$; for $u \ll 0$ (e.g., region $n_0 > 0$) $\epsilon_{10}$ is small and the transition does not occur until much larger $B$. While quantitatively predicting the location of the transition requires accounting for some additional interaction effects (e.g. the Lamb shift and inter-layer capacitance\cite{Hunt16}), such a transition is clearly seen in our experiment. Fig.~\ref{fig:full_fan} shows that for $u > 0$, there is a transition at $\nu = -4 + 2$ around $B \sim 17$T ( indicated by the white arrow).
This is  the transition between filling $\ket{\uparrow, 0},  \ket{\uparrow, 1}$ (low $B$) and filling $\ket{\uparrow, 0}, \ket{\downarrow, 0}$ (high $B$).
No analogous transition is observed for $u < 0$ at $\nu = 2$, at least up to $B = 44$T.

The analysis, then, can be summarized as follows. For the $n_0 < 0, u > 0$ side of the experiment, at high $B$ there is a  large splitting $\epsilon_{10}$ between $\ket{+ \sigma 0}$ and $\ket{+ \sigma 1}$, and the levels fill in order $\ket{+ \uparrow, 0}, \ket{+ \downarrow, 0}, \ket{+ \uparrow, 1}, \ket{+ \downarrow, 1}$ for $-4 < \nu < 0$.
In contrast, for the $n_0 > 0, u < 0$ side of the experiment, the splitting $\epsilon_{10}$ between $\ket{+\sigma 0}$ and $\ket{+\sigma 1}$ is much smaller, and the levels fill in order $\ket{+ \uparrow, 0}, \ket{+ \uparrow, 1}, \ket{+ \downarrow, 0}, \ket{+ \downarrow, 1}$ for $0 < \nu < 4$, at least in the absence of a moir\'e potential. The moir\'e will ``mix'' the $N=0, 1$ LLs in this regime, as we will see.

\begin{figure*}[ht!]
\begin{center}
\includegraphics[width=6in]{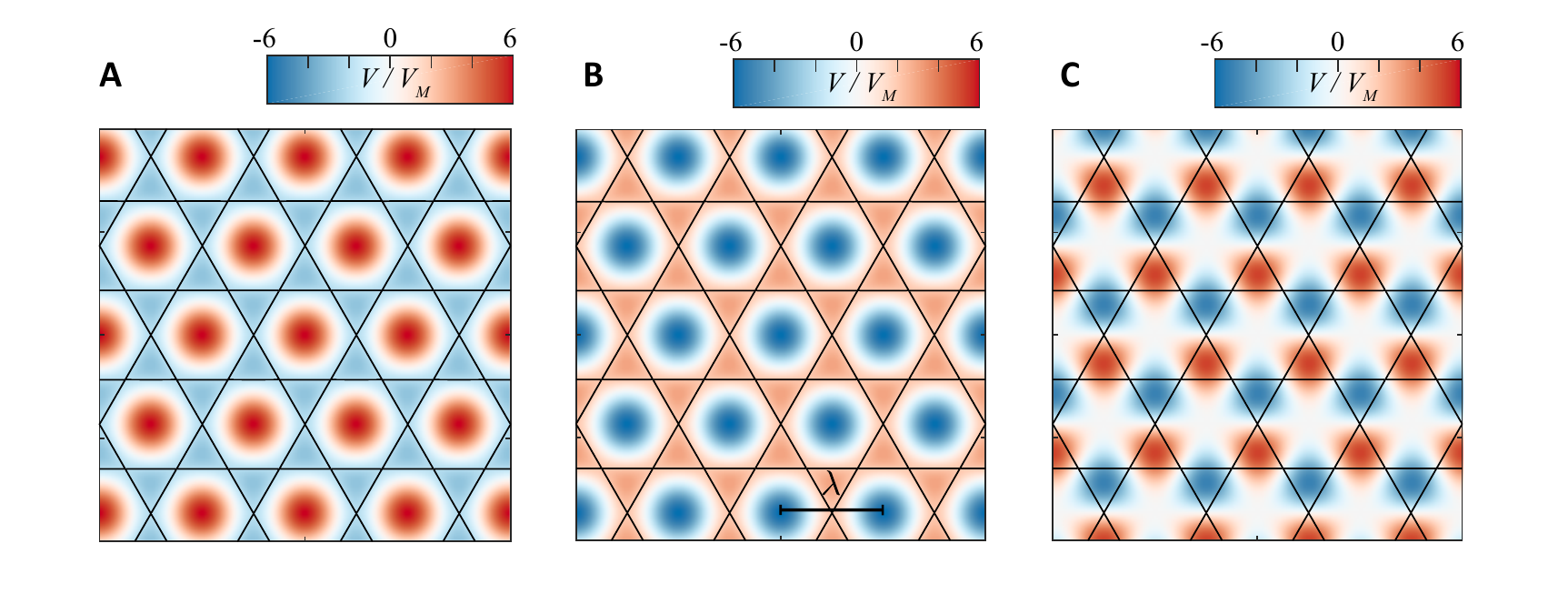} 
\caption{
\textbf{Calculated moir\'e potentials.}
 Real-space moir\'e potentials for \textbf{(A)} $V_M$ = 1, or $|V_M| =$ 1, $\Theta_M$ = 0, \textbf{(B)} $V_M$ = -1 or $|V_M| =$ 1, $\Theta_M = \pi/3$  \textbf{(C)} $V_M$ = $i$ or $|V_M| =$ 1, $\Theta_M = \pi/6$. The postive (negative) $V_M$ potential is repulsive (attractive)   on the triangular lattice, and attractive (repulsive) on the honeycomb lattice.
\label{fig:moire}}
 \end{center}
 \end{figure*}
 
\subsection{Effect of the moir\'e potential on the ZLL}

	Following the existing literature, and the absence of Hofstadter features in states localized on the bottom layer, we assume that the moir\'e pattern affects only the top layer of the BLG, leading to a six-parameter phenomenological model whose effective two-band Hamiltonian for BLG is given in Ref.~\cite{chen_zero-energy_2016}. Because the amplitude $V_M$ of the moir\'e is small compared to $\hbar \omega_c$, we project the moir\'e Hamiltonian into the ZLL. This assumption is supported by the experimental observation that the cyclotron gaps at $\nu  = \cdots -8, -4, 4, 8 \cdots$ remain robust up to  $n_\phi \sim 1$, which implies the moir\'e potential is weak compared to the cyclotron energy. The effective moir\'e Hamiltonian~\cite{chen_zero-energy_2016} simplifies drastically when projected into the ZLL, consisting of only a scalar potential. The simplest form of the moir\'e which is $C_3$ symmetric is of the form
\begin{equation}
V_M(\vec{r}) = V_M \sum_{m=0, 1, 2}  e^{i \mathbf{G}_m \cdot \vec{r}} + \textrm{h.c.}, \quad V_M = |V_M| e^{i \theta_M}. \label{eq:moire}
\end{equation}
Here $\mathbf{G}_m = \hat R_{\frac{2 \pi m }{3}} \mathbf{G}_0$ are the minimal reciprocal vectors of the moir\'e pattern. 

Taking $\theta_M \to \theta_M + \frac{2 \pi}{3}$ leaves the model invariant up to a translation, while under inversion $\theta_M \to - \theta_M$.
We note three special cases: (a) $\theta_M = 0$ ($V_M > 0$): an inversion-symmetric triangular lattice in which sites are repulsive (b) $\theta_M = \frac{2 \pi}{6}$ (equivalent to $V_M < 0$): an inversion-symmetric triangular lattice in which sites are attractive (c) $\theta_M = \pm \frac{\pi }{6}$ ($V_M = \pm i$): an inversion anti-symmetric lattice. 
Cases (a-c) are shown in Fig.~\ref{fig:moire}.
Microscopically, there is no inversion symmetry, and no consensus exists on the more realistic choice of $\theta_M$.

\subsection{Hamiltonian of minimal model} 
	Projecting into the $+$-valley of the ZLL, a minimal model for the system is then
\begin{equation}
\begin{split}
H = \frac{1}{2} \int d^2 q \, \, \rho_{\textrm{ZLL}+}(-\vec{q}) V_C(q) \rho_{\textrm{ZLL}+}(\vec{q}) + \\ \int d^2 r  V_M(\vec{r}) \rho_{\textrm{ZLL}+}(\vec{r}) + \epsilon_{10}\hat{N}_{N=1} + E^{(Z)} \sigma^z \\
\label{eq:H}
\end{split}
\end{equation}
where $\rho_{\textrm{ZLL}+}(\vec{q})$ is the projected 2D density operator, which requires the use of BLG ``form factors'' as reviewed in Ref.~\cite{Hunt16}.
The model includes (1) a spin-SU(2) symmetric Coulomb interaction; (2) a moir\'e  potential parameterized by complex amplitude $V_M$  (Eq.~\eqref{eq:moire}); (3) a splitting $\epsilon_{10}$ between the $N=1$ and $N=0$ orbitals, which depends on $B$ and $u$ (4) a Zeeman splitting. We ignore the small SU(4)-breaking valley interaction anisotropies; they are unlikely to play a role here due to the large valley splitting $u$.

\subsection{Single particle analysis}
	
    We begin with a single-particle analysis to compare the non-interacting (integer $t,s$) features we observe to the expected Hofstadter spectrum. Given $E_C > V_M$, interactions may change the observed Hofstadter spectrum significantly, and we do not necessarily expect quantitative agreement with experiment.
    
	While the limit in which the lattice potential is the largest scale has received the most attention of late, leading to a tight-binding problem with complex hopping amplitudes, \cite{Harper55,azbel_energy_1964,Hofstadter76,haldane_model_1988} in our experiment the lattice is weak compared with the cyclotron gap. 
In this limit is is appropriate to consider the Hamiltonian of Eq.~\eqref{eq:H}, where the lattice potential is projected into the continuum Landau levels, as analyzed in Refs.~\onlinecite{Langbein69, PfannkucheGerhardts92}.  Hints of this physics were observed in semiconducting quantum wells with patterned superlattice. \cite{Gerhardts91, Schlosser96}

\begin{figure*}
\includegraphics[width = 5 in]{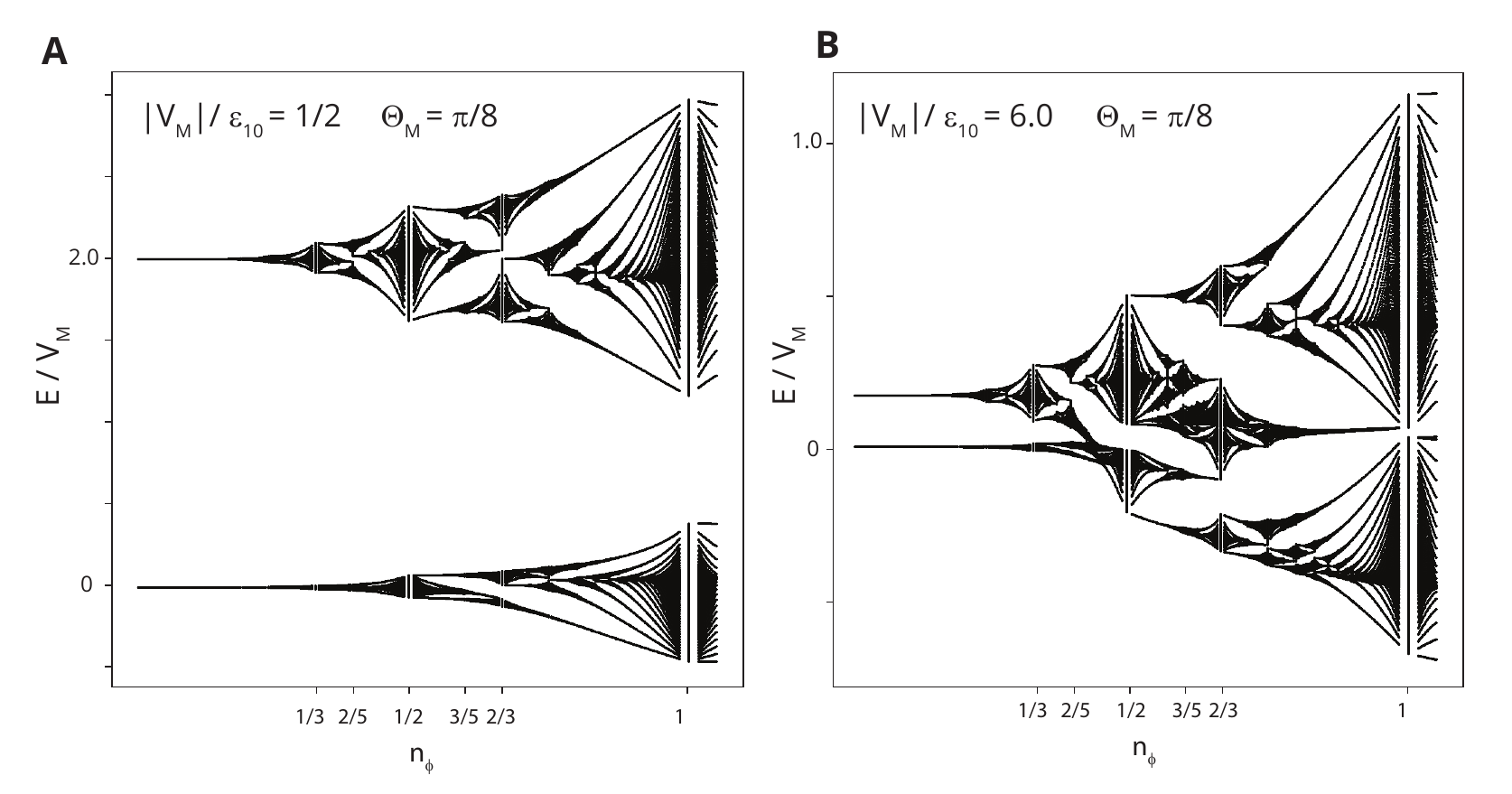}
\caption{
\textbf{Single particle Hofstadter butterfly.}
Single-particle Hofstadter spectrum calculated using the moir\'e parameters \textbf{(A)} $|V_M|/\epsilon_{10} = 1/2.0$ and $\Theta_M = \pi /8$ and \textbf{(B)} $|V_M|/\epsilon_{10} = 6.0$ and $\Theta_M = \pi /8$.
\label{fig:Hof_LR}}
\end{figure*}

 At the single-particle level, the two spins decouple and the phase diagram depends only on the complex ratio $V_M / \epsilon_{10}$. Two illustrative Hofstadter spectra are shown in Fig.~\ref{fig:Hof_LR}.
At low $n_\phi$, the energy spectrum collapses into two flat bands separated by $\epsilon_{10}$; these are  the $N=0, 1$ continuum LLs. This is consistent with experiment, where Hofstadter features only begin appearing around $n_\phi > 1/3$.
This can be qualitatively understood because potentials which vary faster than $\ell_B$ are invisible to the low LLs.  Quantitatively, when potential $V_M(\vec{G})$ is projected into LL $N=0  , 1$, it is scaled by the factor $\mathcal{F}_{00}(G) = e^{-(G \ell_B)^2 / 4}$ and  $\mathcal{F}_{11}(G) = e^{-(G \ell_B)^2 / 4} (1 - (G \ell_B)^2/2 )$ respectively. As $(G \ell_B)^2 / 4 = \frac{\pi}{\sqrt{3} n_\phi}$, the  potential vanishes at low $n_\phi$. 
The $N=1$ level also develops bandwidth faster than the $N=0$ level, because of the factor $(1 - (G \ell_B)^2/2 ) < -1$ in the effective $N=1$ potential.  

Interestingly, when $|V_M| / \epsilon_{10} \gtrapprox 1$, bands with different Chern number can be realized for different values of $\epsilon_{10}$, which can be tuned with $u$ (e.g. $p_0$) in our measurements.
This is evident in the differences in single-particle gaps which appear in our measurements at positive and negative $p_0$ (Figs.~1C-F).
In the future, this could be used to engineer the butterfly spectrum in-situ. 
To compare this model with experiment, we analyze the $u>0$ and $u<0$ cases separately, as they have very different $|V_M|$/$\epsilon_{10}$ ratios. We are able to recover many of the single-particle features we observe by fine tuning  $|V_M|$/$\epsilon_{10}$ and $\Theta_M$ of the model moir\'e pattern.
	
\subsubsection{Case I: $n_0 < 0, u > 0$.} 

In this regime, several observations are consistent with our assertion that $u$ leads to a large splitting $\epsilon_{10}$ between the $N=0, 1$ orbitals.  For large $\epsilon_{10}$, we expect the filling order is  $\ket{+ \uparrow, 0}, \ket{+ \downarrow, 0}, \ket{+ \uparrow, 1}, \ket{+ \downarrow, 1}$. This  order is supported by the presence of a feature at $\nu = -3$, indicated by an arrow in Fig.\ref{fig:full_fan}, which presumably marks a phase transition from the previously reported \cite{Hunt16} lower-$u$ filling order($N = \ket{0 \uparrow}, \ket{1 \uparrow} , \ket{0 \downarrow}, \ket{1 \downarrow}$).

The $|V_M|$/$\epsilon_{10} < 1$ limit is also consistent with several other experimental observations: (1) the LL gaps at $\nu = -4, -3, -2, -1$   persist across $n_\phi = 1/2$, indicating $V_M$ is too weak to overcome $\epsilon_{10}$ at this magnetic field;  (2) filling $-4 < \nu < -3$ looks  similar to $-3 < \nu < -2$ (both are dominated by $\mathcal{C} = -1, 2$ Chern-bands, with FCIs in the first $\mathcal{C} = -1$ band), while  $-2 < \nu < -1$ looks more similar to $-1 < \nu < 0$. This again supports the filling order $\sigma N = \uparrow 0, \downarrow 0, \uparrow 1, \downarrow 1$.
(c) The Hofstadter features begin appearing at $n_\phi \sim 1/2$ for filling $-4 < \nu < -2$, while they appear earlier, around $n_\phi \sim \frac{1}{3}$, for $-2 < \nu < 0$. This is consistent with the expected broader moir\'e induced bandwidth of the $N=1$ levels, which fill after the $N=0$ orbitals.

\begin{figure*}
\includegraphics[width = 6 in]{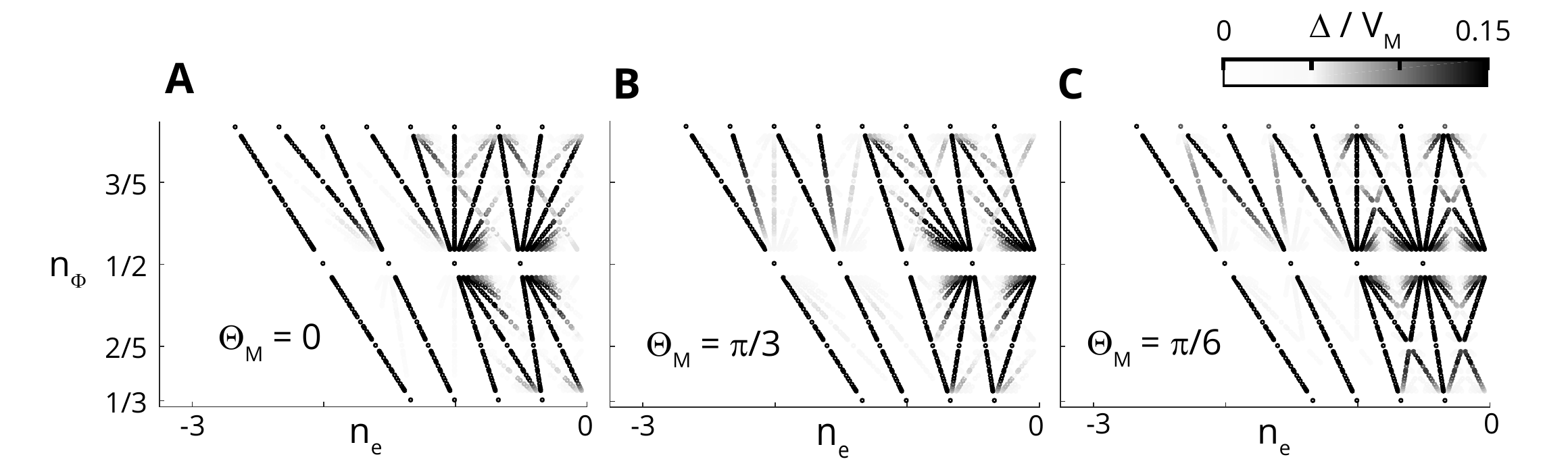}
\caption{
\textbf{Calculated Wannier plots for the $n_0<$0, $u>$0 case.}
Calculated gap sizes in units of the magnitude of the moir\'e potential ($\Delta / V_M$) as a function of $n_e$ and $n_\Phi$ for $\Theta_M$ = 0 \textbf{(A)}, $\Theta_M = \pi/3$ \textbf{(B)}, and $\Theta_M = \pi/6$  \textbf{(C)}. The splitting between LLs $\epsilon_{10}$ is chosen to be larger than any Hofstadter gaps. 
\label{fig:wannier_case1}}
\end{figure*}

To compare our single particle model with experiment, we calculated the single-particle Hofstadter butterfly for $V_M$ with different $\Theta_M$ in the limit of $V_M/\epsilon_{10} \rightarrow 0$ (the result is qualitatively unchanged for small but finite $V_M/\epsilon_{10}$). In Fig.~\ref{fig:wannier_case1}, we plot the calculated single particle gaps on a Wannier plot for the three cases shown in Fig.~\label{Vmoire}, assuming the $N = \ket{0}, \ket{0} , \ket{1}, \ket{1}$  filling order described previously. In the $N=0$ orbital, $\Delta t = -1,2$ bands are prominent in the data, and theoretically are predicted for $\Theta_M = 0, \pi/6$ but not $\pi/3$. The inversion-odd case ($\pi / 6$, Fig.~\ref{fig:wannier_case1}C) favors features which are more particle-hole symmetric within a LL, and lead to crossing features at low $n_\Phi$ which are observed in the data (Fig.~1C). Comparing with our data in this regime, we thus conclude that $\Theta_M$ is somewhere between 0 and $\pi/6$. 

\begin{figure*}
\includegraphics[page = 1,width = 3 in]{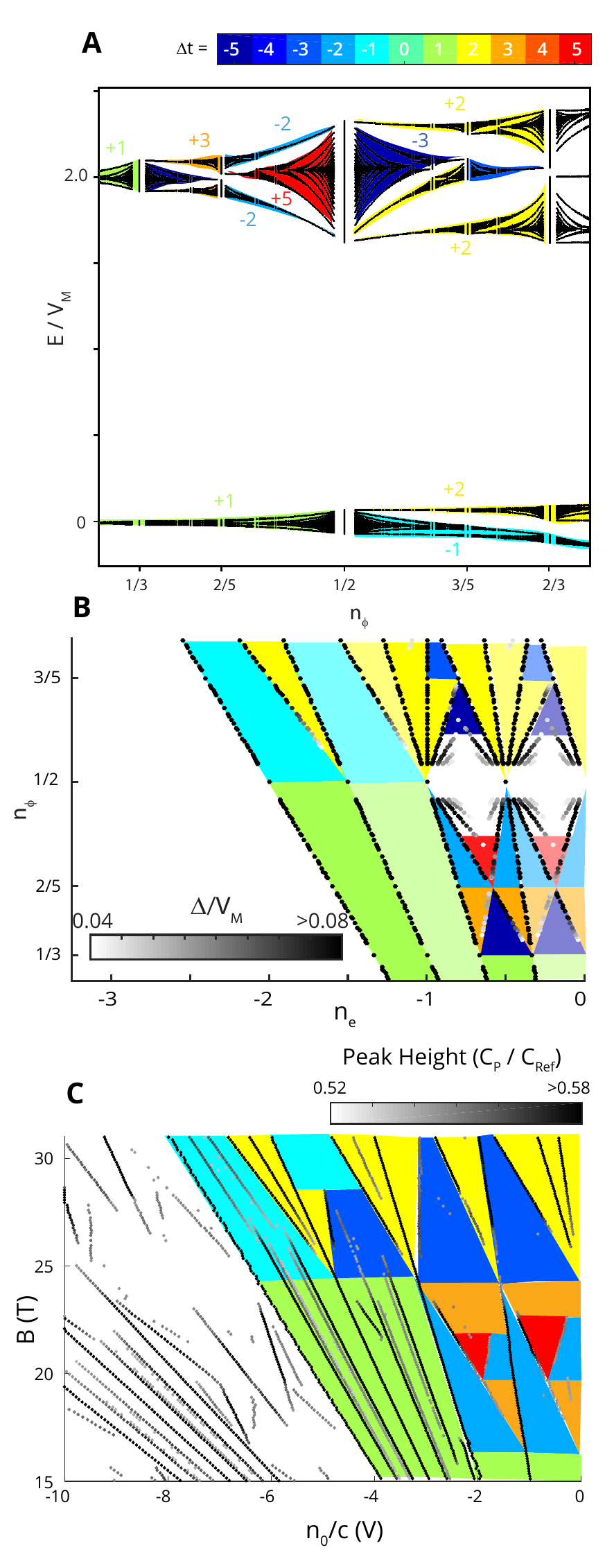}
\caption{
\textbf{Comparison of calculated and observed single particle Chern bands.}
\textbf{(A)} Single-particle Hofstadter spectrum calculated using the moir\'e parameters $V_M/\epsilon_{10} = 2.0$ and $\Theta_M = \pi /8$, chosen to match the $n_0 < 0$, $u > 0$ case. $\Delta t = \mathcal{C}$ bands are labeled from the following Wannier plots using the procedure described in the main text (Fig.~1G).
\textbf{(B)} Calculated Wannier plot constructed from (A). The points are colored according to the size of the gap ($\Delta / V_M$), while the bands are colored according to Chern number. To match the experimental data, the spectrum in (A) was separated by orbital and tiled with filling order $N = 0, 0, 1, 1$, and we set the filling factor to start at $\nu = -4$. The non-transparent bands match the spectrum calculated in (A). 
\textbf{(C)} Peak height of gapped states (black to gray points), extracted from data in Fig.~1C, as a function of charge carrier density ($n_0/c$) and magnetic filed ($B$). The bands are colored according to their single-particle $\Delta t$, using the same rules as (B).
\label{fig:coloredButterfly_L}}
\end{figure*}

By tuning $\Theta_M = \pi / 8$, we observe good agreement between the calculated Chern band structure and the observed bands (Fig.\ref{fig:coloredButterfly_L}). We used the calculated Hofstadter spectrum (Fig.\ref{fig:coloredButterfly_L}A) to generate a Wannier plot (Fig.\ref{fig:coloredButterfly_L}B) which matches the filling order of orbitals observed in the data. The color of the points encodes the size of the single particle gap ($\Delta/|V_M|$), and we only plot gaps above a threshold, in effect cutting off the fractal nature of the Hofstadter spectrum. We color the bands of the energy spectrum and Wannier plot based on their Chern numbers, using the rules outlined in the main text. 

In Fig.\ref{fig:coloredButterfly_L}C we generate a qualitatively equivalent plot from the data (from Fig.~1C) by plotting the height of peaks in $C_P$ as a function of density ($n_0/c$) and magnetic field ($B$). We again color the Chern bands, now ignoring gapped trajectories which we know to be outside of the single-particle picture.

In $N=0$ orbitals, we observe $\Delta t = -1, 2$ bands filling above $n_\phi = 1/2$ in both the calculation and experiment. Additionally, the appearance of $\Delta t = -2,3,5$ bands below $n_\phi = 1/2$ in $N=1$ bands is consistent. Some features, e.g the differences in filling order of Chern bands in  $N=1$ orbitals above $n_\phi = 1/2$, cannot be reproduced without invoking mixing between spin species, which is not allowed in the single particle model we present.

\paragraph{Details of $\mathcal{C} = -1$ band hosting FCI states.} The $\nu_C = 1/3, 2/3$ FCI states discussed in the main text occur in a $\mathcal{C} = -1$ band of the $n_0 > 0, u < 0$ region, which, as discussed above, arises in a model of a moir\'e potential with $0 < \Theta_M < \pi/6$ projected into a single $\ket{+0 \sigma}$ level. In our DMRG calculations (see next section) we choose $\Theta_M = \pi/8$, which reproduces most of the large single-particle gaps in the $n_0 < 0, u > 0$ side of the data as well.

\begin{figure*}
        \includegraphics[width=4in]{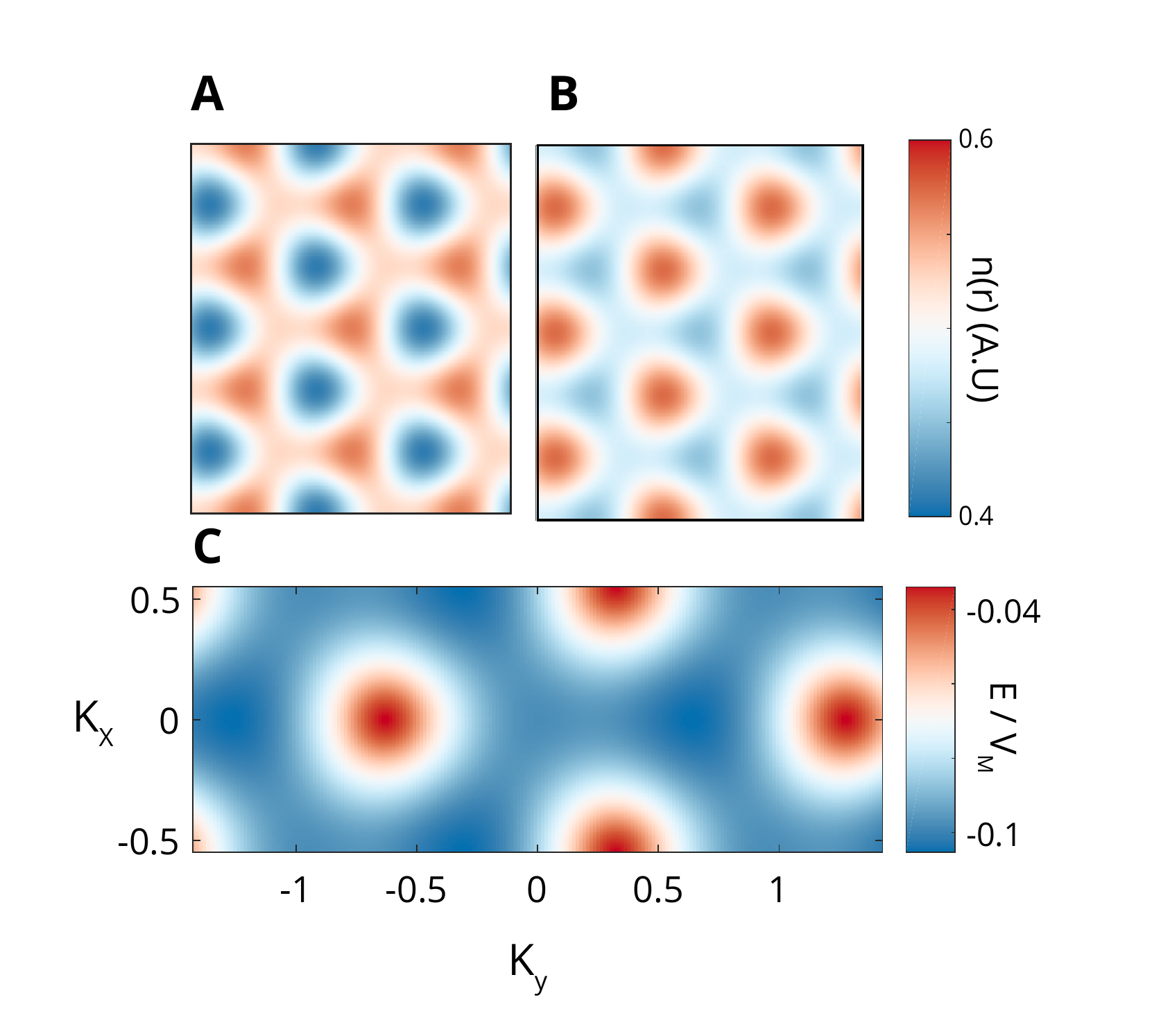}
    \caption{\textbf{Real-space electron density and k-space dispersion of FCI-hosting $\mathcal{C}=-1$ band.} For large $\epsilon_{10}/|V_M|$  at  $n_\phi = \frac{2}{3}$, the $N=0$ level splits up into $\mathcal{C} = -1, 2$ bands. In (\textbf{A,B}), we show the real-space charge density that arises from fully filling each of these bands. We see that the $\mathcal{C} = -1$ band is localized on the triangular lattice. This is the band which hosts the observed FCI states. In \textbf{(C)}, we show the $k$-space energy dispersion of the $\mathcal{C} = -1$ band, with axes in units of $\ell_B^{-1}$ and energy in units of $|V_M|$. At $n_\phi = \frac{2}{3}$, the bandwidth is about 10\% of $|V_M|$; this percentage grows as $n_\phi \to 1/2$. \label{fig:FCIband}}
\end{figure*}

In Fig.~\ref{fig:FCIband}, we show the real-space charge density profile and energy dispersion of this band at $n_\phi = \frac{2}{3}$, assuming $V_M/\epsilon_{10} = 1/6$ and $\Theta_M = \pi/8$.
From the density profile, we see that the $\mathcal{C} = -1$ band is localized on a triangular lattice. 

	\subsubsection{Case II: $n_0 > 0, u < 0$.} In this regime, we expect $\epsilon_{10}$ is smaller and, in the absence of a moir\'e potential, the levels would fill in order $\ket{+ \uparrow, 0}$, $\ket{+ \uparrow, 1}$,  $\ket{+ \downarrow, 0}$, $\ket{+ \downarrow, 1}$. This is consistent with experiment: the $\nu = 2$ LL gap is robust,  and the $0 < \nu < 2$  physics looks very similar to the $2 < \nu < 4$ physics. Strikingly, we see that the $\nu = 1, 3$ LL gaps are destroyed near $n_\phi = 1/2$. For $n_\phi = \frac{1}{2}, 0 < \nu < 2$,  the system has rearranged from $\mathcal{C} = 1, 1$ LLs into $\mathcal{C} = -1, 3$ Chern-bands, which requires $V_M$ comparable to $\epsilon_{10}$.

\begin{figure*}
\includegraphics[width = 6 in]{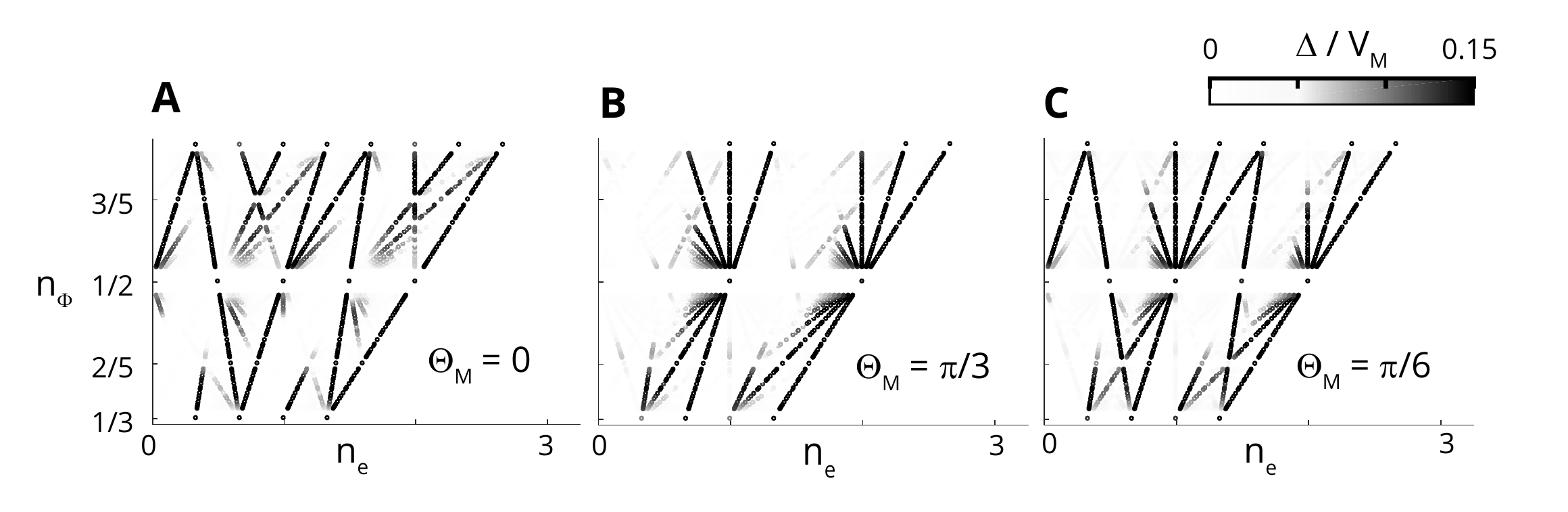}
\caption{
\textbf{Calculated Wannier plots for the $n_0>$0, $u<$0 case.}
Calculated gap sizes in units of the magnitude of the moir\'e potential ($\Delta / V_M$) as a function of $n_e$ and $n_\Phi$ for $\Theta_M$ = 0 \textbf{(A)}, $\Theta_M = \pi/3$ \textbf{(B)}, and $\Theta_M = \pi/6$  \textbf{(C)}. $|V_M|/\epsilon_{10}$ is fixed to 6.0 for all cases. The spin-split gap at $\nu = 2$, and the cyclotron gap at $\nu = 4$, is chosen to be larger than any Hofstadter gaps.
\label{fig:wannier_case2}}
\end{figure*}

The calculated Wannier plots with $|V_M| / \epsilon_{10}$ = 6.0 shows the dependence on the moir\'e parameter $\Theta_M$, where the strength of the moir\'e now strongly mixes $N=0$ and $N=1$ orbitals of the same spin (Fig.~\ref{fig:wannier_case2}). 

At the single particle level, we can ask which values of $|V_M| / \epsilon_{10}$ and $\Theta_M$ rearrange  filling $0 < \nu < 2$ at $n_\phi = 1/2$   into $\mathcal{C} = -1, 3$ bands. We find that $\Theta_M = 0, \pi/6$ both reproduce this behavior, while $\Theta_M = \pi/3$ does not. Weaker features, such as the presence of $\mathcal{C} = 5$ bands around $n_\Phi = 2/5$ favor a more antisymmetric form of the moir\'e potential (e.g. $\Theta_M$ close to $\pi/6$). The value $\pi / 8$ used in our DMRG numerics satisfies these constraints.

\begin{figure*}
\includegraphics[page = 2,width = 3 in]{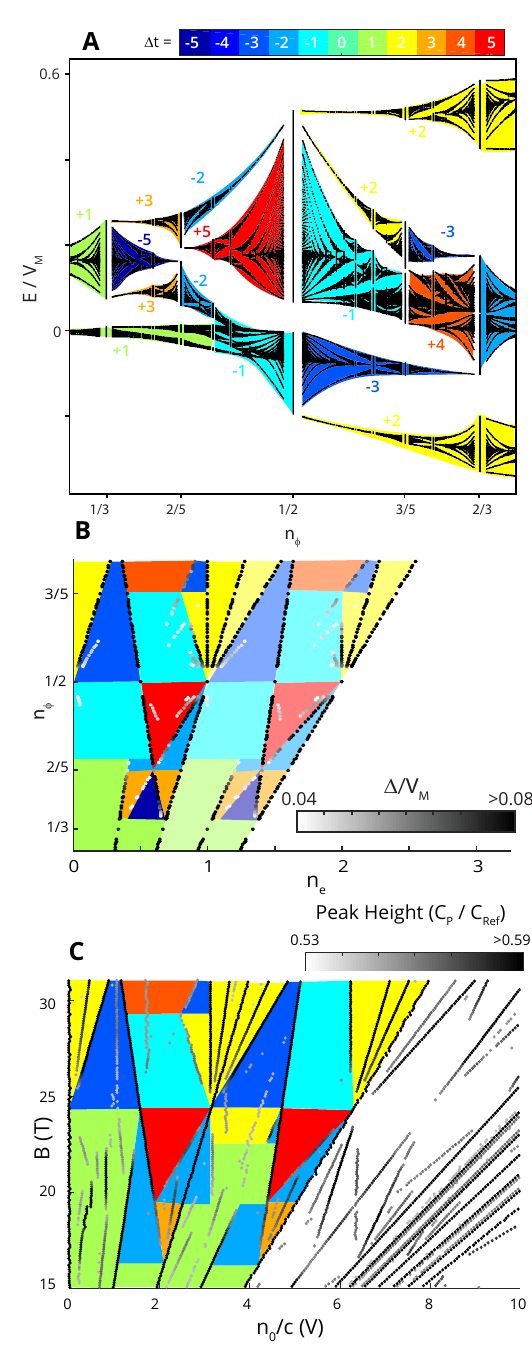}
\caption{
\textbf{Comparison of calculated and observed single particle Chern bands for case II.}
\textbf{(A)} Single particle Hofstadter spectrum using $|V_M|/\epsilon_{10} = 1/6$ and $\Theta_M = 0.5$
\textbf{(B)} Calculated Wannier plot constructed from (D). The whole spectrum was replicated twice to match the experimental data, with the non-transparent bands matching the states represented in (D). 
\textbf{(C)} Peak height of gapped states (black to gray points), extracted from data in Fig.~1D. The bands are colored according to their single-particle $\Delta t$, using the same rules as (E).
\label{fig:coloredButterfly_R}}
\end{figure*}

	As before, tuning  $\Theta_M$, generates good agreement between the calculated Chern band structure and the observed bands (Fig.\ref{fig:coloredButterfly_R}). We find that $\Theta_M =0.5$, $V_M/\epsilon_{10} = 6.0$ gives slightly better agreement for some weaker gaps above $n_\phi = 1/2$ than $\Theta_M =\pi/8$ (which is used in the iDMRG calculations), but the $\Delta t = +3$ band where we performed the calculation is robust in both cases. 
    We color both the calculated and experimental Wannier plots (Fig.\ref{fig:coloredButterfly_R}B-C) in the same way as before. Here, we replicated the entire mixed Hofstadter spectrum to reproduce $2 <\nu <4$ and a large  $\nu = 2$ gap is assumed. This matches the observed filling order of $\ket{+ \uparrow, 0}, \ket{+ \uparrow, 1},  \ket{+ \downarrow, 0}, \ket{+ \downarrow, 1}$.
    Focusing on $0 > \nu > 2$ in the experimental data, we find a very close match between the observed Chern bands. Many features of the data are reproduced, including the $n_\phi$ onset of Hofstadter features in $N=0$ and $N=1$ orbitals, disappearance of the $\nu = 1$ gap at $n_\phi=1/2$, presence of a $\Delta t$ = 5 band above $n_\phi=2/5$ and the first and last filled $\Delta t = +2$ bands above $n_\phi$ =1/2. 
    Deviations between the theory and experiment are primarily in smaller gap features. For example, the calculated spectrum shows low field $\Delta t$ = +3,-5 bands, which are a single $\Delta t$ = -2 band in the data. Small adjustments to the moir\'e potential, disorder, or indeed the same interactions which lead to  FCI and SBCI physics could all in principle change the sizes of smaller gaps enough to generate these discrepancies.

\section{infinite DMRG simulations}

Here we present infinite DMRG simulations of the model just derived.
Following our discussion, the simulations are \emph{not} simulations of a tight-binding lattice model, rather, we project the interactions and lattice potential into the continuum LLs of the ZLL.
While a number of numerical works have considered fractional quantum Hall physics in the opposite Harper-Hofstadter tight-binding limit, \cite{PhysRevLett.94.086803,  PhysRevLett.96.180407,PhysRevA.76.023613, PhysRevLett.103.105303,Neupert2011, Regnault2011, PhysRevB.86.165314, PhysRevB.90.115132, PhysRevB.90.075104, Sheng2011, Neupert2011, Regnault2011, LeeThomaleQi2013, PhysRevB.93.235133}
less attention has been payed to the weak-potential limit of the present experiment. \cite{PfannkucheMacDonald97}

	We will consider both an FCI and SBCI, in  both cases choosing a moir\'e parameter $\Theta_{M} = \frac{\pi}{8} \sim 0.4$ which (at the single particle level) is consistent with all the dominant integer CI features.

\subsection{$\nu_C = \frac{1}{3}$ FCI in $\mathcal{C} = -1$ band.}
The $\mathcal{C} = -1$ band detailed in Fig.~2 of the main text can be accounted for if $|V_M| / \epsilon_{10}$ is small, as discussed above. Since the simulations are challenging in the presence of the moir\'e potential, and $V_M / \epsilon_{10}$ is small, we make a further approximation by discarding the $\ket{+1\uparrow}$ level, projecting entirely into $\ket{+0\uparrow}$. Following our earlier discussion, we consider the Hamiltonian
\begin{align}
H &= \frac{1}{2} \int d^2 q \, \, \rho(-\vec{q}) V_{RPA}(q) \rho(\vec{q}) +   \int d^2 r  V_M(\vec{r}) \rho(\vec{r}) \\
V_M(\vec{r}) &= V_M \sum_{m=0, 1, 2}  e^{i \mathbf{G}_m \cdot \vec{r}} + \textrm{h.c.}
\end{align}
where $\rho(q)$ is the density operator projected into a single $N=0$ LL. The Coulomb interaction is $V_C(q) = E_C \frac{2 \pi}{q} \tanh(q d /2 )$ (where $q$ is in units of $\ell_B^{-1}$), due to screening from the graphite gates at a distance $d \sim 10 \ell_B$ from the BLG. Here $E_C = \frac{e^2}{\epsilon \ell_B}$ is the Coulomb scale.
However, having projected out the other LLs,  to make more quantitative comparison with experiment we also include RPA screening from the filled LLs below the ZLL, \cite{PapicAbanin}

\begin{equation}
\begin{split}
V_{RPA}(q) = \frac{V_C(q)}{1 + V_C(q) \Pi(q) },\\ \quad \Pi(q) = a 4 \log(4) \tanh(b q^2 \ell_B^2) / 2 \pi E_C
\label{eq:HiDMRG}
\end{split}
\end{equation}
The screening weakens the short-distance part of the Coulomb interaction. 
While not essential to the existence of the FCI - we also find the FCI state without it - it does change the range of $|V_M|/E_C$ where the FCI is stabilized by around $\sim 20\%$, since it effectively reduces the Coulomb scale.
Following comparison between DMRG numerics and experimental data in an earlier work,\cite{Hunt16} we take $a = 0.2 \frac{E_C}{\hbar \omega_c}, b = 0.62$ where $\hbar \omega_c$ is the cyclotron energy at the desired field.
For the moir\'e, we choose $V_M = e^{2 \pi i / 16} |V_M|$ (this choice of $\Theta_M$ reproduces the experimentally observed CIs in our measurements), while $|V_M|/E_C$ is a tunable parameter to be explored.

\begin{figure*}
\includegraphics[width = 6in]{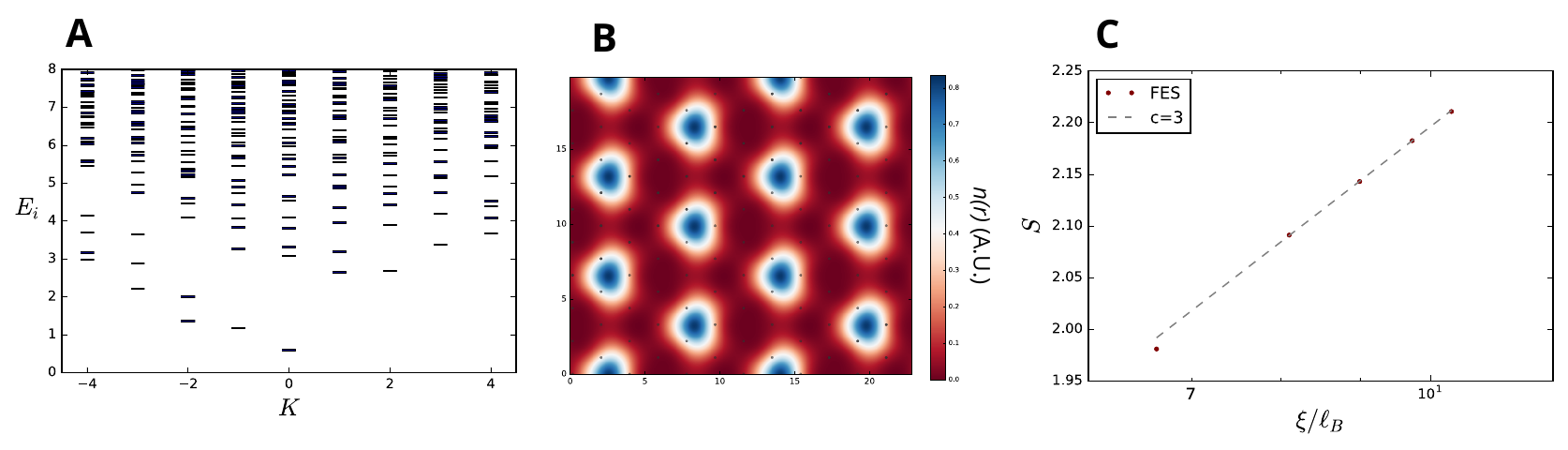}
\caption{ \textbf{Entanglement spectrum of $\nu_C$ = 1/3 FCI and investigation of competing phase.s}
\textbf{(A)} Entanglement spectrum of the $C=-1, \nu_C = \frac{1}{3}$ FCI. The low-lying counting disperses from right to left as 1, 1, 2, 3, (5), where it merges into the higher states. Note that in our convention, the $\nu = \frac{1}{3}$ Laughlin state would have counting $1, 1, 2, 3, 5, \cdots$ with the \emph{opposite} chirality, left to right. This reversal is a signature of the reversed Hall conductance in a $C = -1$ band. 
\textbf{(B)} Charge carrier density $\langle n(r) \rangle$ at $\nu_C = \frac{1}{3}$ filling of the $C=-1$ band in the regime where $|V_M| / E_C < 0.29$ leads to a Wigner crystal. The dots indicate the moir\'e unit cell, showing a 3x3 reconstruction.
\textbf{(C)} Evidence for a gapless phase at $\nu_C = \frac{1}{3}$ filling of the $C=-1$ band in the regime  $0.79 < |V_M| / E_C$. We measure the evolution of the bipartite entanglement entropy $S$ vs. DMRG correlation length $\xi$ as the DMRG bond dimension increases. In a CFT, $S = \frac{c}{6} \log(\xi) + s_0$. At the point $|V_M| / E_C = 2$ shown here, we obtain perfect agreement with $c=3$. At higher $|V_M|$ (not shown) we find $c=6$.
}
\label{fig:FCInumerics}
\end{figure*}

	iDMRG proceeds by placing the above continuum quantum Hall problem  onto an infinitely long cylinder of circumference $L$.\cite{ZaletelMongPollmannRezayi}
iDMRG requires an ordering of the single-particle states into a 1D chain, which arises naturally on the cylinder when the LL orbitals are taken in the Landau gauge. 
We emphasize again that the ``sites'' in our chain are not the minima of the moir\'e potential, but rather the orbitals of the continuum LL.
To accommodate the triangular moir\'e lattice with Bravais vectors $\vec{a}_1, \vec{a}_2$, we form a cylinder by identifying $\vec{r} \sim \vec{r} + 9 \vec{a}_1$.
Working at $n_\phi = \frac{2}{3}$, this corresponds to a cylinder of circumference $L \approx 19.8 \ell_B$.
$\nu_C = \frac{1}{3}$ filling of the $\mathcal{C} = -1$ band corresponds to $\nu = \frac{1}{6}$ of the $N=0$ LL.
iDMRG\cite{mcculloch_infinite_2008} using $m = 3000$ states was used to find the ground state for a range $|V_M| / E_C$.
The lattice reduces the continuous translation symmetry of the cylinder down to $\mathbb{Z}_9$, making the simulations  more expensive; nevertheless, the DMRG truncation error was less than $3 \cdot 10^{-6}$ throughout the FCI phase.
 	
	For an intermediate range of $0.29 < |V_M| / E_C < 0.74$, we find a state with a short correlation length ($\xi \sim 3 \lambda$, where $\lambda$ is the period of the moir\'e lattice) and  $t,  s = -\frac{1}{3}, \frac{1}{3}$ (we ignore electrons below the ZLL), which we thus identify as an FCI.
The entanglement spectrum of the FCI is shown in Fig.~\ref{fig:FCInumerics}A, and is consistent with a Laughlin type state but with negative Hall conductance \cite{LiHaldane}.

We measure $(t, s)$ as follows.
Since $\nu =  t +  s / n_\phi$ (tautologically), it is sufficient to measure either $s$ or $t$, and in our simulations it is  most convenient to measure $s$.
We do so by repeating iDMRG for a series of moir\'e potentials which are displaced by a distance $\Delta x$ along the cylinder, $V(\vec{r}) = V_M( \vec{r} - \Delta x)$, obtaining a sequence of ground states $\ket{\Delta x}$.
By definition, $s$ is the amount of charge per unit cell which should be transported along with the lattice. The charge which passes a cut around the cylinder is  $\Delta Q =  s L \Delta x / \mathcal{A}$, where $\mathcal{A}$ is the volume of the unit cell.
We can measure the amount of charge transported $\Delta Q$ by using the entanglement spectrum to compute the charge polarization of $\ket{\Delta x}$, as discussed for an analogous measurement of the Hall current in Ref.~\onlinecite{zaletel_flux_2014}. To ensure adiabaticity, $\Delta x$ was incremented in units of $\ell_B / 24$ using the previous ground state to initialize the DMRG. The results give a  perfectly quantized value for $s$ within the $10^{-6}$ precision of the numerics.

For $|V_M| / E_C < 0.29$, the ground state is found to increase the unit cell with a 3x3 reconstruction, forming a triangular Wigner crystal shown in Fig.~\ref{fig:FCInumerics}B. Effectively, all the electrons in the ZLL are inert: $t = 0, s = \frac{1}{9}$. This is to be expected, since the Coulomb interaction alone stabilizes a Wigner crystal at such low fillings ($\nu = \frac{1}{9}$).
The location of the transition can be diagnosed from $\langle \rho(\mathbf{G}_0 / 3) \rangle$, where $\mathbf{G}_0$ is a reciprocal vector of the moir\'e. 
To see the symmetry breaking, the numerics must be done with an enlarged unit cell and lower degree of momentum conservation.

For $ 0.74 < |V_M| / E_C$, there is a change in the correlation length and entanglement properties as the system  enters a compressible phase through what appears to be a continuous phase transition.
This region is rather complex.
When $|V_M| / E_C \to \infty$, the system should be a non-interacting metal due to the small but finite bandwidth of the Chern band.
It is very  interesting question whether, in 2D, there is a direct transition between the FCI and this metal, or whether an intermediate state (such as a composite Fermi liquid or symmetry broken phase) intervenes.
However, this 2D physics is subtle to address on the cylinder, where we suspect there is in fact a sequence of several KT-transitions.
To see this, we used ``finite entanglement scaling'' \cite{PollmannFES}
to measure the central charge $c$ of the cylinder state. At $|V_M|/E_C = 2$, we find a very precise value of $c=3$ (Fig.~\ref{fig:FCInumerics}C), while at $|V_M|/E_C = 6$ we find $c=6$.
Multiples of 3 are expected, because the magnetic algebra at $n_\phi = \frac{2}{3}$ enforces a 3-fold degeneracy in the Fermi surface.
The 2D Fermi surface of the metal descends to a several-component Luttinger liquid on the cylinder due to the quantization of the momentum around the cylinder. As $|V_M| / E_C $ changes the Luttinger exponents, it naturally could drive a sequence of KT-transitions at which some, but not all, of the modes lock.
In precisely the same regime that finite entanglement scaling finds a finite central charge, we also observe a weak ``stripe''-like order; translation is preserved along $a_1$, but $a_2$ is broken. This can be diagnosed from $\langle \rho(\mathbf{G}'/2) \rangle$ for an appropriate reciprocal vector. It is difficult to determine whether this is a true property of the ground state, or is instead a finite-entanglement artifact, $\langle \rho(\mathbf{G}'/2) \rangle \propto \xi_{\textrm{FES}}^{-\Delta}$, where $\xi_{\textrm{FES}}$ is a correlation length introduced by the finite bond dimension of our DMRG numerics.
Regardless, it gives a very clear indication of the onset of the gapless phase, so is the metric we presented in the phase diagram of the main text.

	For comparison with experiment, we note that $E_C = 48$meV at $B=32$T ($n_\phi = \frac{2}{3}$) assuming a dielectric constant of $\epsilon = 6.6$ for the surrounding BN. This gives the estimate $14 < |V_M| < 38$meV for an FCI, consistent with the expected moir\'e amplitude.

\subsection{$\nu_C = \frac{p}{3}$ SBCI in $\mathcal{C} = 3$ band.}

	The $\mathcal{C} = 3$ band hosting the SBCI state detailed in the main text emanates from $\nu, n_\phi = 2, \frac{1}{3}$.
As discussed, near $n_\phi = \frac{1}{2}$ the stability of this $\mathcal{C} = 3$ band requires a small $\epsilon_{10}$ which mixes the $N=0, 1$ levels.
However, we have verified that near $n_\phi = \frac{1}{3}$, the $\mathcal{C}=3$ band remains stable even as $\epsilon_{10} \to \infty$.
In this limit, the $N=0$ level is completely filled and inert, and the potential $V_M$ is effectively projected into an $N=1$ level.
While a quantitative study of the SBCI may require keeping both $N=0, 1$ levels and finite $\epsilon_{10}$, this is numerically challenging, so we take advantage of this finding to take $\epsilon_{10} \to \infty$ and project the problem into the $N=1$ level. The Hamiltonian is the same as in Eq.~\ref{eq:HiDMRG}, but now $\rho(q)$ is the density operator projected into a $N=1$ LL (in fact if we incorrectly project into $N=0$ level, we do not find an SBCI). We take $V_M = |V_M| e^{2 \pi i / 16}$ as before.

We again place the problem on the cylinder, but this time we identify $\vec{r} \sim \vec{r} + 12 \vec{a}_1 $. This was chosen to accommodate a tripled unit cell with enlarged Bravais vector  $\vec{a}_1 + \vec{a}_2$.
We work at $n_\phi = \frac{3}{8}$, where $\nu_C = \frac{1}{3}, \frac{2}{3}$ of the $\mathcal{C} = 3$ band correspond to filling $\nu = 1 + \frac{7}{9}$ and $\nu = 1 + \frac{8}{9}$ (the integer part of the filling is now assumed to occupy an inert $N=0$ LL). iDMRG was performed while keeping 3000 states. We have not obtained a full phase diagram for $|V_M| / E_C$, but found a range of  values (e.g. $|V_M| / E_C = 0.6$ in the main text) which stabilize an SBCI state and are consistent with the domain of the $C = -1$ FCI.
The SBCI is diagnosed by a tripled unit cell (seen in the real-space density) and the experimentally predicted $t, s$, again measured by adiabatically dragging the lattice.
We note that working on an \emph{infinitely} long cylinder greatly simplifies the detection of the symmetry breaking. Because the symmetry is discrete, it can be spontaneously broken in this geometry, unlike in finite-size simulation on a torus.

\bibliography{refs}
\clearpage
\section {Additional experimental data}

\begin{figure*}[ht!]
\begin{center}
\includegraphics[width=6.5in]{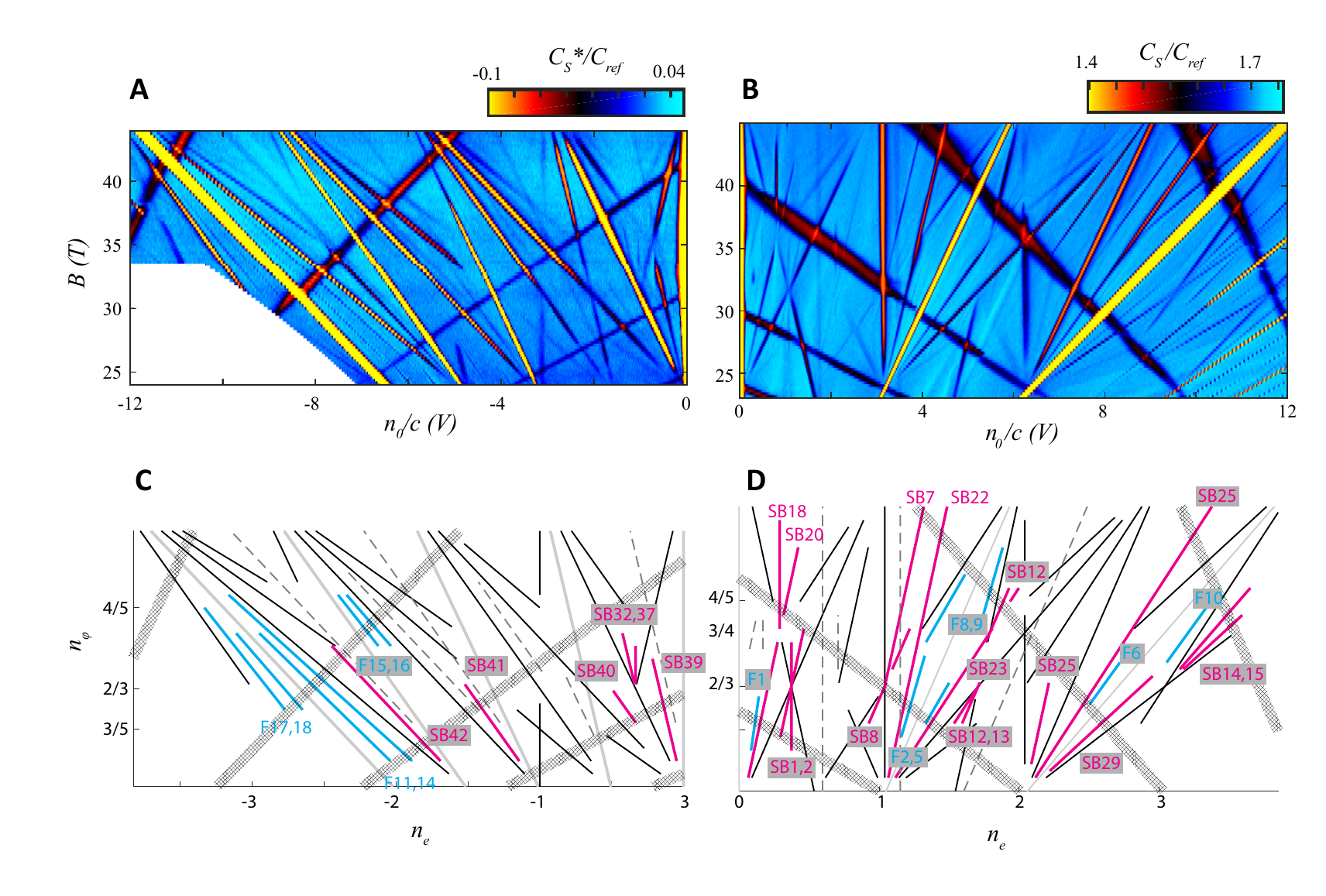} 
\caption{\textbf{Data from a measurement run up to 45T.}
\textbf{(A,B)} Landau fans of $C_S$ for $p_0/c$ = 16 V (A) and - 16 V (B). We performed a horizontal line-by-line subtraction on (A). \textbf{(C,D)}  
Classification of gapped linear trajectories (A) and (B) respectively. Interaction-driven features are labeled and their $t,s$ are given in Tables S1 and S2.
We observe two classes of linear trajectories which do not fall into the categories outlined in the main text. The first are associated with Landau Levels in one of the graphite gates, which appear much more prominently in $C_S$ than in $C_P$ and only depend on one of the applied gate voltages (hatched wide lines). These states are observed as diagonal features in the $n_0-p_0$ plane, and appear as a secondary, broad Landau fan with an x-intercept that depends on $p_0/c$. The second are features that either do not have $t$ and/or $s$ which clearly match a small-denominator rational fraction, are short-lived in $B$, or do not have nearby features which allow us to easily identify their origin (dashed lines). The FCI states described in the main text persist between 27 and nearly 40 T. 
\label{fig:hybrid}}
\end{center}
\end{figure*}

\begin{figure*}[hb!]
\begin{center}
\includegraphics[width=4in]{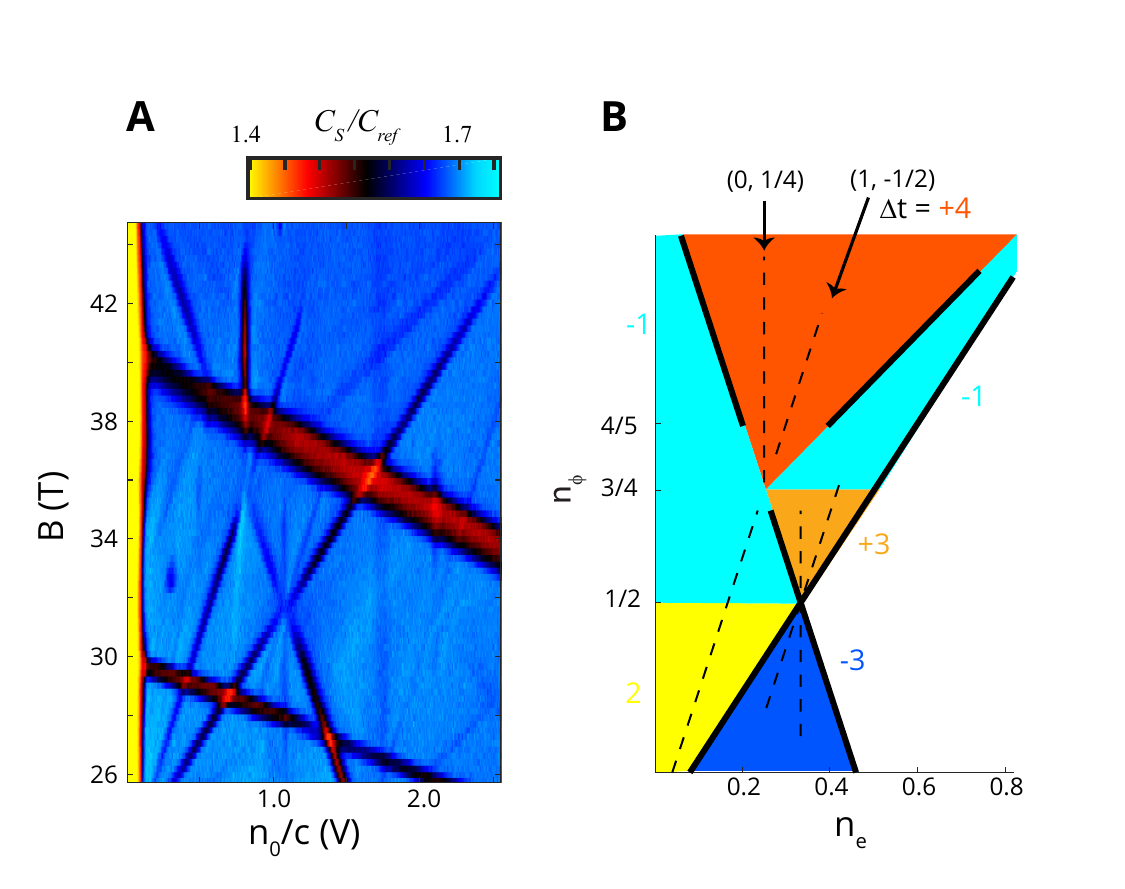} 
\caption{\textbf{Symmetry broken Chern insulators in a $C$=4 band.}
\textbf{(A)} Detail from Fig.\ref{fig:hybrid}, highlighting a $C$ = $\delta t$ = 4 band at high magnetic field. Broad, negative s. \textbf{(B)} Schematic of (A), SBCI states (dashed lines) with $(t,s) = (0,1/4)$ and $(1,-1/2)$ occur at $\nu_c$ = 1/4 and 2/4 fractional filling of a $\Delta t = 4$ band (dark orange). SBCI in $\Delta t = 2$ and $\Delta t = \pm3$ are also observed. Broad, negative slope features not represented in the schematic are due to graphite LLs. \label{sbci2}}
 \end{center}
 \end{figure*}

\begin{figure*}[ht!]
\begin{center}
\includegraphics[width=6in]{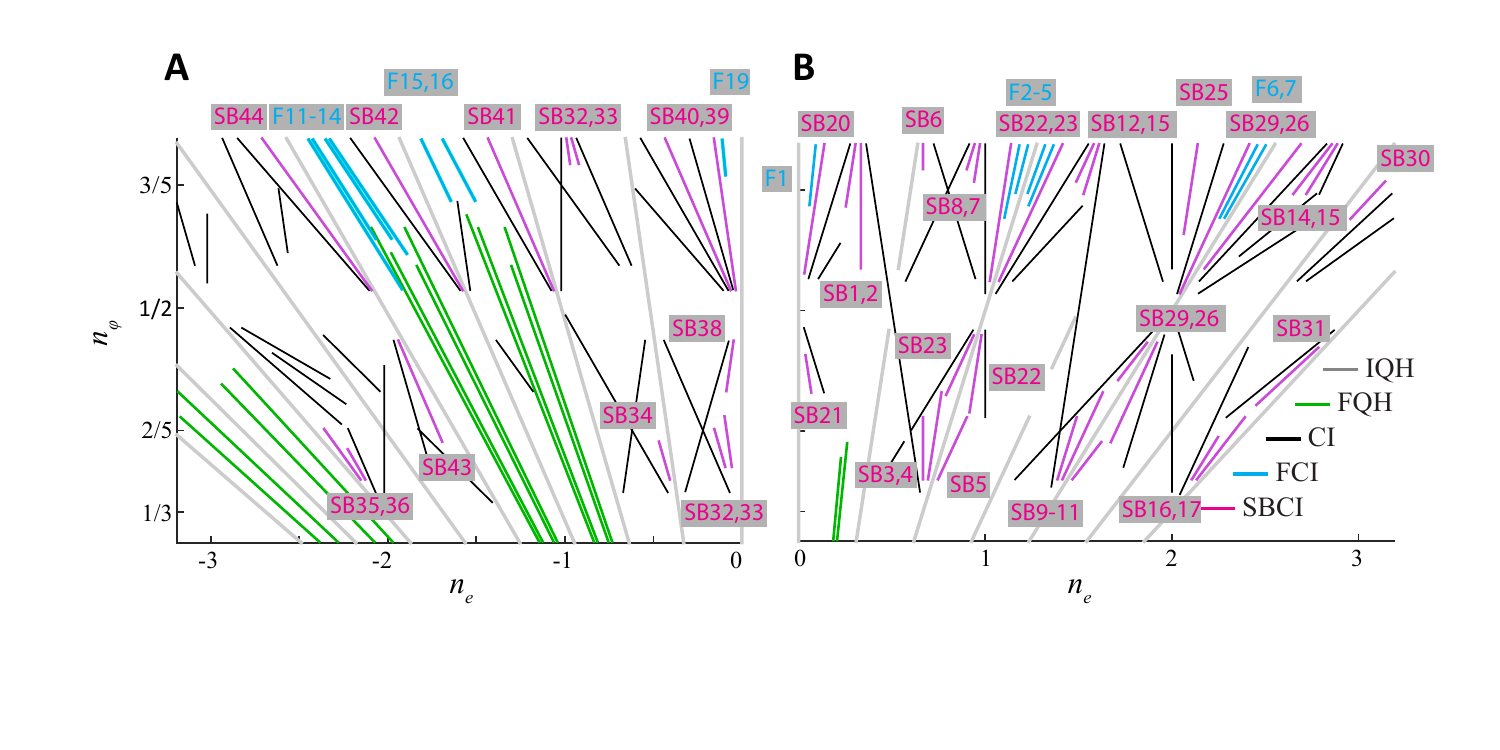} 
\caption{\textbf{Labeled gap trajectories for Fig.~1E-F.}
Annotated version of Fig.~1 with interaction driven states labeled according to Tables \ref{SBCI_table} and \ref{FCI_table}. 
\label{annotated}}
\end{center}
\end{figure*}

\begin{wraptable}{l}{3.4 in}
\caption{\textbf{List of observed symmetry-broken Chern insulator states (integer $t$, fractional $s$).} States above 30 T can be observed in Fig.\ref{fig:hybrid}. 'id's refer to labels in Fig.\ref{fig:hybrid} and Fig.\ref{annotated}}
\label{SBCI_table}
\centering
\begin{tabular}{ccccc}
id & $t$ & $s$ & B [T] {(}min,max{)}&$p_0/c$ [V]\\
\hline \hline
   SB1& 1  & -1/3 & (28,36)   & -16\\
   SB2& 0 &  1/3 &  (26,35)   & -16\\
   SB3&  0 & 2/3& (17,20)   & -16 \\
   SB4&  1&  1/3&   (17,21)    & -16\\
   SB5&  3&  -1/3&  (17,20) & -16\\
   SB6&   0&  2/3&  (30,31)    & -16\\   
   SB7&  1&  1/3&  (30,45)    & -16\\   
   SB8&  2&  -1/3&  (30,36)    & -16\\  
   SB9&  2&  2/3&  (17,20)    & -16\\  
   SB10&  3&  1/3&  (17,21)    & -16\\   
   SB11&  5&  -1/3&  (17,19)    & -16\\  
   SB12&   3&  -1/3&  (29,32)    & -16\\   
   SB13&  2&  1/3&  (29,32)    & -16\\   
   SB14&  5&  -1/3&  (29,31) (33,37)   & -16\\  
   SB15&  4&  1/3&  (29,39)    & -16\\  
   SB16&  5&  1/3&  (17,20)    & -16\\   
   SB17&  4&  2/3&  (17,19)    & -16\\  
   SB18&  0&  1/4&  (36,44)    & -16\\
   SB19&  2&  1/4&  (35,39)    & -16\\
   
   SB20&  1& -1/2&  (25.8,35) (37, 42) & -16\\  
   SB21& -1&  1/2&  (21.5,22.5)    & -16\\   
   SB22&  1&  1/2&  (20.2,23.3) (25.5,45)    & -16\\  
   SB23&  3& -1/2&  (20.9,23.3) (25.5,36.0)    & -16\\  
   SB25&  1&  3/2&  (27.4,32)    & -16\\   
   SB26&  3&  1/2&  (17,23) (25.0,45.0)    & -16\\  
   SB27&  5& -1/2&  (26,32.5)    & -16\\
   SB28&  3&  1/2&  (19,23)    & -16\\  
   SB29&  5& -1/2&  (21.5,23) (25.5,32.5)    & -16\\   
   SB30&  6& -1/2&  (28.0,29.5)    & -16\\  
   SB31&  7& -1/2&  (20.5,22.8)    & -16\\

   \hline
   SB32& -1  & 1/3  & (18,20) (29,36) & 16\\
   SB33& -2 &  2/3 &  (18,20) (29,32) & 16\\
   SB34&  -2&  1/3 & (17,19) & 16 \\
   SB35&  -5&   -1/3&  (18,20) & 16\\
   SB36&  -4& -2/3& (18,19) & 16\\
   SB37&  0& -1/3& (32,35) & 16 \\
   
   SB38& 1 &  -1/2 &  (21,23) & 16\\
   SB39&  -1&  1/2 & (25,32) & 16 \\
   SB40&  -3&   3/2&  (25,31) & 16\\
   SB41&  -3&   1/2&  (25,32) & 16\\
   SB42&  -4&   1/2&  (25,35) & 16\\
   SB43&  -3& -1/2& (19,23) & 16\\
   SB44&  -5& 1/2& (25,32) & 16\\
\end{tabular}
\end{wraptable}

\begin{wraptable}{r}{3.4 in}
\caption{\textbf{List of observed fractional Chern insulator states (fractional $t$, fractional $s$).} * in the magnetic field indicates a lower bound on the field at which states disappear, as these weak states were not clearly observed in the higher field data, possibly due to worse signal to noise and resolution. States above 30 T can be observed in Fig.\ref{fig:hybrid}. 'id's refer to labels in Fig.\ref{fig:hybrid} and Fig.\ref{annotated}.}
\label{FCI_table}
\centering
\begin{tabular}{ccccc}
id & $t$ & $s$ & B [T] {(}min,max{)}&$p_0/c$ [V]\\
\hline \hline
   F1& 2/3  & -1/3  & (28,32)   & -16\\
   F2&  4/3&   1/3&  (28,39)    & -16\\
   F3&  5/3&  1/6 & (29,31*)   & -16 \\ 
   F4&  7/3&  -1/6&  (35,40)    & -16\\ 
   F5&  8/3&  -1/3&  (29,31*) (35,40)    & -16\\ 
   F6&  10/3&   1/3&  (28,39) & -16\\
   F7&  11/3&   1/6&  (28,31*)    & -16\\   
   F8&  8/3&  -2/3& (35,40) & -16\\
   F9&  4/3& 2/3& (36,42)& -16\\
   F10&  10/3& 2/3& (36,42)& -16\\
   
   \hline
   F11& -13/3  & 1/3  & (25,36) & 16\\
   F12& -22/5 &  2/5 &  (27,32*) & 16\\
   F13&  -23/5&  3/5 & (27,32*) & 16 \\
   F14&  -14/3&   2/3&  (26,38) & 16\\
   F15&  -11/3&   2/3&  (28,32*) (35,39) & 16\\
   F16&  -10/3&   1/3&  (28,32*) (35,39)& 16\\
   F17&  -11/3& -1/3& (30,36) & 16\\
   F18&  -10/3& -2/3& (30,38)& 16\\
   F19&   -2/3&  1/3& (29.5,31.5)&16\\
\end{tabular}
\end{wraptable}

\end{document}